%% file: main.tex
\documentclass{article}

\PassOptionsToPackage{numbers,sort&compress}{natbib}
\usepackage[eandd, final]{neurips_2026}
\usepackage{float}
\usepackage[utf8]{inputenc} 
\usepackage[T1]{fontenc}    
\usepackage{hyperref}       
\usepackage{url}            
\usepackage{booktabs}       
\usepackage{amsfonts}       
\usepackage{nicefrac}       
\usepackage{microtype}      
\usepackage{xcolor}         
\usepackage{pifont}
\usepackage{booktabs}
\usepackage{tabularx}
\usepackage{multirow}
\usepackage{graphicx}
\usepackage{wasysym}
\input{macro}

\usepackage{fontawesome5}
\definecolor{vcred}{HTML}{C9222B}     
\definecolor{vcorange}{HTML}{F6A231}  
\definecolor{vclink}{HTML}{A51C24}    
\definecolor{vccite}{HTML}{A85A0C}    
\hypersetup{
  colorlinks=true,
  linkcolor=vclink,
  citecolor=vccite,
  urlcolor=vclink,
  pdftitle={VeriContest: A Competitive-Programming Benchmark for Verifiable Code Generation},
}
\newcommand{\vclogo}[1][0.55in]{\includegraphics[height=#1]{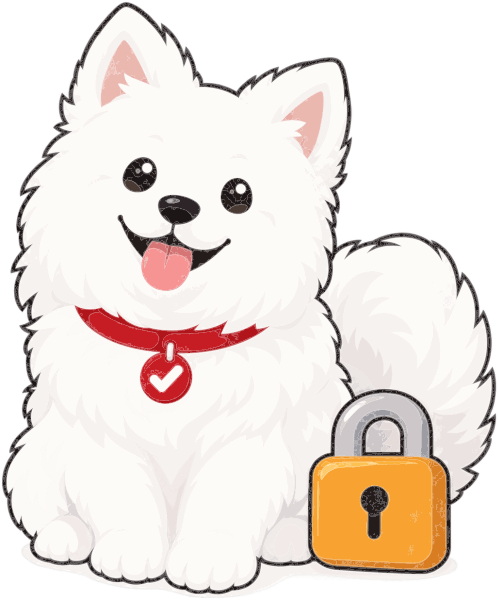}}

\title{\texorpdfstring{%
  $\vcenter{\hbox{\vclogo}}$\hspace{0.7em}%
  $\vcenter{\hbox{\begin{tabular}{@{}c@{}}
    \benchmark: A Competitive-Programming\\
    Benchmark for Verifiable Code Generation
  \end{tabular}}}$%
}{VeriContest: A Competitive-Programming Benchmark for Verifiable Code Generation}}

\author{%
\textbf{Zichen Xie$^{1}$} \quad
\textbf{Mrigank Pawagi$^{2}$} \quad
\textbf{Yuxin Liu$^{3}$} \quad
\textbf{Aaditi Rai$^{1}$} \\
\textbf{Lize Shao$^{1}$} \quad
\textbf{John Berberian Jr.$^{1}$} \quad
\textbf{Sicong Che$^{4}$} \quad
\textbf{Wenxi Wang$^{1}$} \\
$^{1}$University of Virginia \quad
$^{2}$University of Pennsylvania \\
$^{3}$Rice University \quad
$^{4}$Independent Researcher \\
\texttt{\{graysonxie,bzm5zw,zgr3et,ccg3sr,wenxiw\}@virginia.edu} \\
\texttt{mpawagi@upenn.edu \quad ol5@rice.edu \quad sicongche@gmail.com} \\[6pt]
{\color{vclink}\faGlobe}\hspace{0.35em}\href{https://hiprel-group.github.io/VeriContest/}{\textbf{hiprel-group.github.io/VeriContest}}
}

\begin{document}

\maketitle

\begin{abstract}
\input{sections/0-abstract}

\end{abstract}

\input{sections/1-introduction}

\input{sections/2-background}

\input{sections/3-construction}

\input{sections/4-benchmark}

\input{sections/5-evaluation}

\input{sections/6-conclusion}

\section*{Acknowledgments}
We thank Mihajlo Markov, Andrija Marinkovic, Mahadi Hasan and Anamul Hoque Emtiaj for their contributions to identifying specification issues using automated tools and manually validating specifications in our benchmark.

\bibliographystyle{plainnat}
\bibliography{reference}


\appendix

\input{sections/appendix}


\end{document}

%% file: macro.tex
\usepackage[dvipsnames]{xcolor}
\usepackage{minted}
\usepackage{listings}
\usepackage{caption} 
\usepackage{graphicx}
\usepackage{adjustbox} 
\usepackage{subcaption}
\usepackage{wrapfig}
\usepackage{multirow}
\usepackage{enumitem}
\usepackage{xspace}
\usepackage[table]{xcolor}
\usepackage{xcolor}
\usepackage{listings}
\usepackage{tcolorbox}
\usepackage{mdframed}

\DeclareRobustCommand{\benchmark}{VeriContest\xspace}
\DeclareRobustCommand{\benchsize}{1,006\xspace}
\newcommand{\highlight}[2]{{\setlength{\fboxsep}{0.5pt}\colorbox{#1}{#2}}}

\definecolor{codegreen}{rgb}{0,0.6,0}
\definecolor{codegray}{rgb}{0.5,0.5,0.5}
\definecolor{codepurple}{rgb}{0.58,0,0.82}
\definecolor{backcolour}{rgb}{0.97,0.97,0.95}
\definecolor{forestgreen}{rgb}{0.28,0.62,0.37}
\definecolor{codeblue}{rgb}{0,0.5,1}

\newcommand{\CodeIn}[1]{{\small \texttt{#1}}}

\lstdefinestyle{ruststyle}{
    morekeywords={
        as, break, const, continue, crate, else, enum, extern, false, fn, for, if, impl, in, let, loop, match, mod, move, mut, pub, ref, return, self, Self, static, struct, super, trait, true, type, unsafe, use, where, while, dyn, abstract, alignof, become, box, do, final, macro, offsetof, override, priv, proc, pure, sizeof, typeof, unsized, virtual, yield, async, await, try
    },
    sensitive=true, %
    morecomment=[l]{//},  %
    morecomment=[s]{/*}{*/},  %
    morestring=[b]",  %
    morestring=[b]{'}, %
    keywordstyle=\color{codepurple},  %
    commentstyle=\bfseries\color{codeblue}\itshape,  %
    stringstyle=\color{blue},  %
    identifierstyle=\color{black},  %
    ndkeywordstyle=\color{purple}\bfseries,  %
    basicstyle=\ttfamily\scriptsize,  %
    showstringspaces=false,  %
    tabsize=4,  %
    breaklines=true,  %
    breakatwhitespace=false,  %
    showtabs=false,  %
    showspaces=false,  %
    showstringspaces=false,  %
    numbers=left,  %
    numberstyle=\tiny\color{gray},  %
    numbersep=-0.1pt,
    xleftmargin=4pt,
}

\lstdefinelanguage{Verus}{
    style=ruststyle,
    morekeywords=[2]{ requires, ensures, invariant, spec, proof, decreases },
    keywordstyle=[2]\color{red}
}

\definecolor{specbg}{RGB}{255,210,160}   
\definecolor{codebg}{RGB}{220,236,247}   
\definecolor{proofbg}{RGB}{248,222,228}  

\lstdefinestyle{verusblock}{
    language=Verus,
    basicstyle=\ttfamily\scriptsize,
    columns=fullflexible,
    keepspaces=true,
    frame=none,
    framesep=0pt,
    framerule=0pt,
    xleftmargin=2.0em,
    xrightmargin=0pt,
    breaklines=true,
    aboveskip=0pt,
    belowskip=0pt,
    numbers=left,
    numberstyle=\ttfamily\scriptsize,
    numbersep=5pt,
    stepnumber=1,
    firstnumber=last,
}
\lstdefinestyle{verusspec} {style=verusblock, backgroundcolor=\color{specbg}}
\lstdefinestyle{veruscode} {style=verusblock, backgroundcolor=\color{codebg}}
\lstdefinestyle{verusproof}{style=verusblock, backgroundcolor=\color{proofbg}}

\lstdefinestyle{promptstyle}{
    basicstyle=\ttfamily\scriptsize,
    breaklines=true,
    columns=fullflexible,
    keepspaces=true,
    frame=none,
    xleftmargin=0pt,
    xrightmargin=0pt,
    aboveskip=0pt,
    belowskip=0pt,
    showstringspaces=false,
}

%% file: sections/0-abstract.tex
Large language models can generate useful code from natural language, but their outputs come without correctness guarantees. Verifiable code generation offers a path beyond testing by requiring models to produce not only executable code, but also formal specifications and machine-checkable proofs. Progress in this direction, however, is difficult to measure: existing benchmarks are often small, cover only part of the pipeline, lack ground-truth proofs or rigorous specification validation, or target verification settings far from mainstream software development. We present VeriContest, a benchmark of 1,006 competitive-programming problems from LeetCode and Codeforces for verifiable code generation in Rust with Verus. Each problem includes a natural language description, expert-validated formal specifications, judge-accepted Rust code, Verus-checked proofs, and positive and negative test suites. VeriContest is constructed through a three-phase pipeline that scales from manually verified seed problems to semi-automated expansion with human-in-the-loop review. To strengthen benchmark quality, we also use testing to validate postcondition completeness. VeriContest supports both isolated and compositional evaluation of specification, code, and proof generation, and end-to-end verified program synthesis. Evaluating ten state-of-the-art models reveals a sharp gap between ordinary coding and verifiable code generation: the strongest model reaches 97.51\% on natural-language-to-code generation, but only 57.16\% on specification generation, 13.82\% on proof generation, and 6.76\% end-to-end. These results identify proof and specification generation as the central bottlenecks for current models and establish VeriContest as a rigorous platform for evaluating and training future systems that generate code with machine-checkable correctness.

%% file: sections/1-introduction.tex
\section{Introduction}

Large language models (LLMs) can now generate useful code from natural language~\cite{huynh2025large, jimenez2023swe, jain2024livecodebench}, and coding agents are widely used in software development~\cite{agarwal2026ai}. However, generated code still comes without correctness guarantees and may contain functional errors~\cite{wang2025towards} or security defects~\cite{schreiber2025security}. Testing can expose some failures, but it only samples program behavior~\cite{dahl1972structured, xia2025beyond, yang2025knighter}. Formal verification offers a stronger alternative by checking code against a formal specification with a machine-checkable proof~\cite{klein2009sel4, leroy2009formal}. This motivates \emph{verifiable code generation}, where a model must produce not only executable code, but also formal specifications and proofs that certify the code satisfies the specification~\cite{thakur2025clever, ye2025verina}.

Measuring progress in this setting requires benchmarks that faithfully capture the full pipeline. Such benchmarks should provide high-quality specifications, code, and proofs; support both isolated and compositional evaluation of specification generation (SpecGen), code generation (CodeGen), and proof generation (ProofGen); and cover problems that are challenging yet relevant to mainstream programming. Existing benchmarks often fall short on at least one of these dimensions: they are small~\cite{sun2024clover}, focus on only one part of the pipeline~\cite{yang2025verusage}, lack ground-truth proofs or rigorous specification validation~\cite{ye2025verina}, or target verification settings far from mainstream software development~\cite{thakur2025clever}.

\begin{figure}[t]
    \centering
    \includegraphics[width=\linewidth]{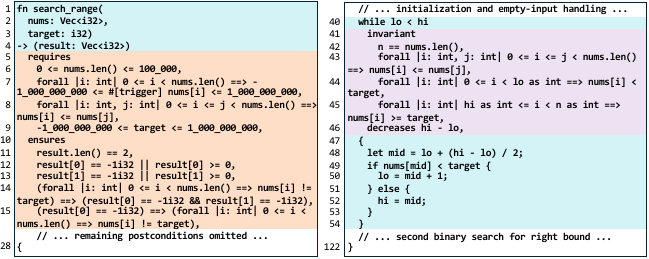}
    \caption{An example Verus problem illustrating the three artifacts of verifiable code generation: \highlight{Orange!25}{specification}, \highlight{SkyBlue!25}{code}, and \highlight{Purple!15}{proofs}. This example (LeetCode 34) returns the first and last indices of the target value (\CodeIn{target}) in a sorted array via two binary searches.}
    \label{fig:verus_example}
    \vspace{-0.5cm}
\end{figure}

Figure~\ref{fig:verus_example} shows how the three components of verifiable code generation appear in a Verus program. The specification includes preconditions (\CodeIn{requires}) and postconditions (\CodeIn{ensures}), which state the required input conditions and output properties, respectively. The executable Rust code implements the algorithm, and the proofs provide the loop invariants, assertions, and lemma functions needed to verify that the code satisfies the specification. In the example, the two binary searches compute the output range, while the proof connects the loop behavior to the postcondition.

We introduce \benchmark, a benchmark of \benchsize competitive-programming problems from LeetCode~\cite{LeetCode2026} and Codeforces~\cite{Codeforces2026} for verifiable code generation in Rust with Verus~\cite{lattuada2024verus}. Each problem contains a natural language description, expert-validated formal specifications, judge-accepted Rust code, Verus-checked proofs, and positive and negative test suites. These problems require algorithmic reasoning over patterns such as greedy methods, dynamic programming, and sliding windows, making them substantially more proof-heavy than elementary programming exercises.

\benchmark is constructed through a three-phase pipeline. We first manually build 91 verified seed problems with sound and complete specifications. We then expand the benchmark to \benchsize problems through semi-automated generation with human-in-the-loop review and refinement. Finally, we generate positive and negative test cases from the verified programs and use them as an additional quality-assurance layer for validating postcondition completeness. In particular, we design \emph{Post2Exe} to convert supported Verus postconditions into executable programs, which are then run on negative test cases to expose incomplete postconditions at scale.

\benchmark supports both isolated and compositional evaluation across the verifiable code generation pipeline, covering SpecGen, CodeGen, ProofGen, and end-to-end generation. Using this protocol, we evaluate ten state-of-the-art LLMs and find a sharp gap between standard code generation and verifiable code generation. The strongest model, GPT-5.5, reaches 97.51\% on natural-language-to-code generation, but only 57.16\% on specification generation, 13.82\% on proof generation, and 6.76\% end-to-end. These results identify specification and proof generation as central bottlenecks for current models. They also show that end-to-end success requires the generated specification, code, and proof to align on the same problem, driving all models below 7\%.

\textbf{Contributions.} (1) We release \benchmark, a benchmark of \benchsize competitive-programming problems with expert-validated Verus specifications, judge-accepted Rust code, Verus-checked proofs, and comprehensive positive and negative test suites. (2) We design a three-phase construction pipeline combining manually verified seed problems, semi-automated expansion with human-in-the-loop review, and test-based validation of postcondition completeness through Post2Exe. (3) We define a modular evaluation protocol for SpecGen, CodeGen, ProofGen, and end-to-end verified program synthesis, and evaluate ten state-of-the-art LLMs to expose the key bottlenecks in current verifiable code generation.

%% file: sections/2-background.tex
\newcommand{\cmark}{\ding{51}}
\newcommand{\xmark}{\ding{55}}
\newcommand{\codegen}{CodeGen}
\newcommand{\specgen}{SpecGen}
\newcommand{\proofgen}{ProofGen}

\section{Related Work}

Table~\ref{tab:related} compares \benchmark with prior LLM benchmarks for code generation and verification. We discuss these works by their primary evaluation focus: SpecGen, CodeGen, ProofGen, and end-to-end verifiable code generation.

\input{tables/related_work.tex}

\textbf{SpecGen.}
FormalSpecCpp~\cite{chakraborty2025formalspeccpp} and FormalBench~\cite{le2025can} focus on generating specifications for C++ and Java programs, respectively, but isolate SpecGen from code and proof generation. \benchmark instead provides expert-validated Verus specifications with corresponding Rust code and proofs, enabling SpecGen to be evaluated both independently and as part of downstream verifiable code generation.

\textbf{CodeGen.}
HumanEval~\cite{chen2021evaluating} and MBPP~\cite{austin2021program} established standard function-level evaluation, while APPS~\cite{hendrycks2021measuring} and LiveCodeBench~\cite{jain2024livecodebench} move toward harder programming and competitive-programming problems. These benchmarks evaluate code only; \benchmark retains competitive-programming difficulty while adding formal specifications and ground-truth proofs, enabling broader evaluation.

\textbf{ProofGen.}
DafnyBench~\cite{loughridge2024dafnybench} evaluates proof generation in Dafny, miniCodeProps~\cite{lohn2024minicodeprops} in Lean, and VerusBench~\cite{yang2025autoverus}, RepoVBench~\cite{zhong2025rag}, VeruSAGE~\cite{yang2025verusage}, and VeriStruct~\cite{sun2026veristruct} for Rust programs. Dafny-Synthesis~\cite{misu2024towards}, FVAPPS~\cite{dougherty2025proving}, AlgoVeri~\cite{zhao2026algoveri}, and VERICODING~\cite{bursuc2025benchmark} combine code and proof generation under other formal-verification settings. These works expose ProofGen's difficulty, but are not centered on real-world competitive-programming tasks in Rust, where solutions are proof-heavy and require reasoning over greedy, dynamic programming, and sliding-window algorithms. \benchmark combines this setting with ground-truth proofs and modular task definitions, allowing proof generation to be studied both independently and as part of end-to-end verifiable code generation.

\textbf{End-to-end verifiable code generation.}
Clover~\cite{sun2024clover} studies closed-loop consistency among code, specifications, and proofs in Dafny. CLEVER~\cite{thakur2025clever}, VERINA~\cite{ye2025verina}, and VerifyThisBench~\cite{deng2025verifythisbench} broaden end-to-end evaluation in Lean or mixed verification languages. VeriBench~\cite{miranda2025veribench} also targets end-to-end Lean verification, but evaluates translation from Python files to verified Lean programs rather than generation from problem descriptions. These benchmarks are closest to \benchmark because they cover all three core capabilities, but are smaller in scale, target languages other than Rust, use different input settings, or lack ground-truth proofs and full compositional evaluation. \benchmark provides \benchsize competitive-programming tasks in Rust, with ground-truth proofs and task compositions that evaluate SpecGen, CodeGen, ProofGen, and end-to-end verifiable code generation in one benchmark.

%% file: tables/related_work.tex
\begin{table*}[t]
    \small
    \centering
    \caption{A comparison of \benchmark with prior LLM benchmarks for code generation and verification. We characterize each work along the three foundational tasks for verifiable code generation: \specgen, \codegen, and \proofgen; \CIRCLE{}, \LEFTcircle{}, and \Circle{} denote full, partial, and no support, respectively. For benchmarks supporting multiple tasks, we annotate whether they support evaluation in a compositional manner; for those including \proofgen, we indicate whether ground-truth proofs are provided. Distinct from prior works, \benchmark provides large-scale evaluation over \benchsize tasks drawn from real-world competitive programming, targeting Rust as a mainstream programming language.}
    \label{tab:related}
    \setlength{\tabcolsep}{4pt}
    \resizebox{\textwidth}{!}{%
    \begin{tabular}{@{}lccccclr@{}}
        \toprule
        Benchmark & \specgen & \codegen & \proofgen & Compositional & GT Proof & Language & Size \\
        \midrule
        FormalSpecCpp \cite{chakraborty2025formalspeccpp} & \CIRCLE & \Circle    & \Circle   & --     & --      & C++        & 105 \\
        FormalBench \cite{le2025can} & \CIRCLE & \Circle    & \Circle   & --     & --      & Java       & 700 \\
        \midrule
        HumanEval \cite{chen2021evaluating}, MBPP \cite{austin2021program} & \Circle & \CIRCLE    & \Circle   & --     & --      & Python     & 164 / 974 \\
        APPS \cite{hendrycks2021measuring}        & \Circle & \CIRCLE    & \Circle   & --     & --      & Python     & 10{,}000 \\
        LiveCodeBench \cite{jain2024livecodebench} & \Circle & \CIRCLE & \Circle & --     & --      & Python     & 511 \\
        \midrule
        DafnyBench \cite{loughridge2024dafnybench} & \Circle & \Circle    & \CIRCLE   & --     & \cmark  & Dafny      & 782 \\
        Dafny-Synthesis \cite{misu2024towards} & \LEFTcircle & \CIRCLE & \CIRCLE  & \xmark & \xmark  & Dafny      & 153 \\
        miniCodeProps \cite{lohn2024minicodeprops} & \Circle & \Circle    & \CIRCLE   & --     & \xmark  & Lean       & 201 \\
        FVAPPS \cite{dougherty2025proving} & \Circle & \CIRCLE    & \CIRCLE   & \xmark & \xmark  & Lean       & 4{,}715 \\
        AlgoVeri \cite{zhao2026algoveri} & \Circle & \CIRCLE    & \CIRCLE   & \xmark & \cmark  & Mixed      & 77 \\
        VERICODING \cite{bursuc2025benchmark} & \LEFTcircle & \CIRCLE & \CIRCLE  & \xmark & \xmark  & Mixed      & 12{,}504 \\
        Verus-Bench \cite{yang2025autoverus}, RepoVBench \cite{zhong2025rag} & \Circle & \Circle    & \CIRCLE   & --     & \cmark  & Rust       & 150 / 383 \\
        VeruSAGE \cite{yang2025verusage}, VeriStruct \cite{sun2026veristruct} & \Circle & \Circle    & \CIRCLE   & --     & \cmark  & Rust       & 849 / 129 \\
        \midrule
        Clover \cite{sun2024clover} & \CIRCLE & \CIRCLE    & \CIRCLE   & \cmark & \cmark  & Dafny      & 66 \\
        CLEVER \cite{thakur2025clever} & \CIRCLE & \CIRCLE    & \CIRCLE   & \xmark & \xmark  & Lean       & 161 \\
        VERINA \cite{ye2025verina} & \CIRCLE & \CIRCLE    & \CIRCLE   & \cmark & \xmark  & Lean       & 189 \\
        VerifyThisBench \cite{deng2025verifythisbench} & \CIRCLE & \CIRCLE    & \CIRCLE   & \cmark & \cmark  & Mixed      & 154 \\
        \midrule
        \textbf{\benchmark} (ours) & \CIRCLE & \CIRCLE & \CIRCLE  & \cmark & \cmark  & Rust       & \textbf{\benchsize} \\
        \bottomrule
    \end{tabular}%
    }
    \vspace{-2mm}
\end{table*}

%% file: sections/3-construction.tex
\section{Benchmark Construction}

\begin{figure}[t]
    \centering
    \includegraphics[width=0.98\linewidth]{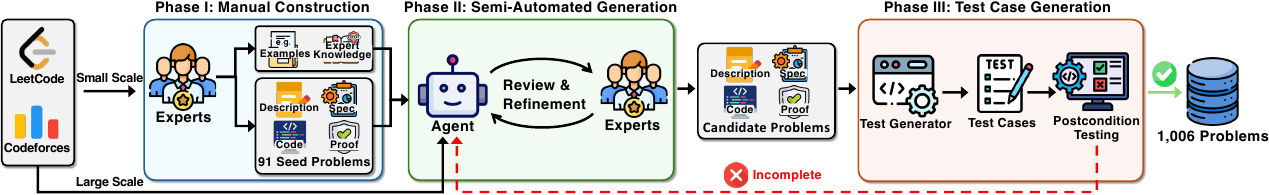}
    \caption{Three-phase construction pipeline for \benchmark.}
    \label{fig:overview}
    \vspace{-0.3cm}
\end{figure}

We curate problems from LeetCode and Codeforces and construct the benchmark in three phases (Figure~\ref{fig:overview}). In \textbf{Phase~I}, we manually write specifications, code, and proofs for 91 seed problems, establishing a high-quality foundation with sound and complete specifications. In \textbf{Phase~II}, we expand the benchmark to \benchsize problems through a semi-automated pipeline with human-in-the-loop review. In \textbf{Phase~III}, we generate positive and negative test cases from the verified programs to further validate postcondition completeness and evaluate model-generated code and specifications. Appendix~\ref{app:benchmark_details} provides additional construction details.

\subsection{Phase~I: Manually Verified Seed Problems}

In Phase~I, we manually construct 91 verified seed problems (81 from LeetCode and 10 from Codeforces). For each problem, we first write the formal specification, then the Rust code, and finally the Verus proof. Each specification must satisfy two core properties with respect to the natural-language problem description:
\begin{itemize}[nosep, left=0pt]
    \item \emph{\textbf{Soundness}}: any code that correctly solves the problem satisfies the specification.
    \item \emph{\textbf{Completeness}}: any code satisfying the specification correctly solves the problem; no invalid solution is admitted.
\end{itemize}
We also avoid specifications that unnecessarily encode a particular implementation strategy, so the benchmark evaluates the intended input--output behavior rather than adherence to one reference algorithm. We validate each seed problem by submitting the Rust code to the online judge and checking the full program with Verus. The seed set serves both as an evaluation subset and as exemplars for the semi-automated expansion in Phase~II.

\subsection{Phase~II: Semi-Automated Verifiable Code Generation}

In Phase~II, we use a coding agent with human-in-the-loop review to expand the benchmark from 91 seed problems to \benchsize problems (722 from LeetCode and 284 from Codeforces). We instantiate the agent with GitHub Copilot, backed by GPT-5.3-Codex, and require the same soundness and completeness criteria used in Phase~I.

\textbf{Problem filtering.} Because Verus supports only a restricted subset of Rust, we exclude tasks requiring unsupported features such as floating-point arithmetic or complex data structures (e.g., priority queues or binary search trees). Before attempting a solution, the agent screens each candidate by checking the problem description and metadata for unsupported features and consulting five top-rated community solutions to determine whether the task can be solved in the supported Rust subset. Problems that fail either check are discarded. Appendix~\ref{app:cr_attrition} reports the filtering and attrition statistics.

\textbf{Expert knowledge and example guidance.} Competitive-programming proofs often require reasoning about algorithmic strategies such as greedy methods, binary search, and sliding windows. To assist the agent, we distill the Phase~I seed problems into reusable lemmas, syntax guidelines, proof templates, and an index of example proofs organized by algorithmic category. These examples and expert knowledge are provided to the agent during construction, allowing it to retrieve relevant proof patterns for each problem.

\textbf{Human-in-the-loop construction and validation.} For each problem that passes filtering, the agent writes the specification from the problem description, uses community solutions to produce correct and efficient Rust code, and generates Verus proofs through iterative repair. We run Verus with the \CodeIn{-{}-no-cheating} flag, which disallows constructs that bypass the verifier: \CodeIn{assume} and \CodeIn{admit} allow properties to be accepted without proof, while \CodeIn{external\_body} and \CodeIn{assume\_specification} cause Verus to trust a function without verifying its body. Human experts then apply three checks: (1) online-judge acceptance for code correctness and efficiency, (2) manual review for specification soundness, completeness, and avoidance of unnecessary implementation-specific constraints, and (3) independent Verus verification for proof validity. Appendix~\ref{app:cr_expert_review} provides the review protocol and validation results.

When the agent fails to verify a problem within 20 minutes, a human expert either provides targeted feedback to the agent or completes the proof manually. Problems that still cannot be verified within an additional 20 minutes are discarded.

\subsection{Phase~III: Test Case Generation}

\begin{figure}[t]
    \centering
    \includegraphics[width=0.98\linewidth]{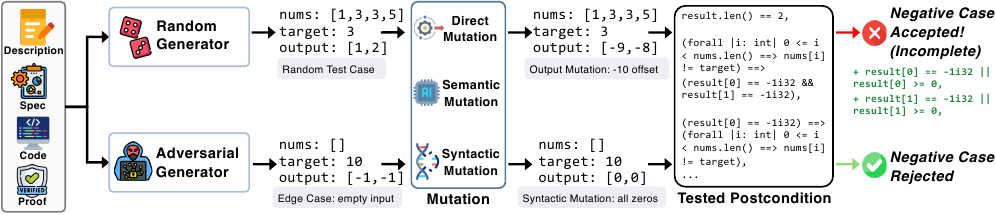}
    \caption{Phase III test-case generation pipeline, illustrated with LeetCode 34. An accepted negative case exposes an incomplete postcondition, which is then strengthened to reject the negative case.}
    \label{fig:post_testing}
    \vspace{-0.3cm}
\end{figure}

Neither LeetCode nor Codeforces publicly releases hidden test cases, yet evaluating generated code and specifications requires concrete inputs and outputs. We therefore derive two complementary test suites from the verified programs (Figure~\ref{fig:post_testing}). \emph{Positive} test cases pair valid inputs with the reference outputs produced by the verified program; \emph{negative} test cases pair valid inputs with incorrect outputs that a complete postcondition should reject. These tests support both benchmark quality assurance and model evaluation (Section~\ref{sec:evaluation_metrics}).

\textbf{Positive test cases.} Following prior work~\cite{wang2025codecontests+, jain2024livecodebench, he2025hardtests, shi2026codehacker, cao2025can}, we generate positive cases with random and adversarial input generators. Rather than asking an LLM to generate test cases directly, we ask it to synthesize a generator program. To guarantee input validity, we use each expert-verified precondition as the generator's postcondition and require the generator to verify in Verus. On average, each problem yields 246.4 positive test cases, achieving 99.60\% line coverage over the verified code.

\textbf{Negative test cases.} We construct negative cases through mutation testing, pairing each valid input with incorrect outputs that should be rejected by a complete specification. We target a negative set ten times larger than the positive set in three stages. First, \emph{semantic mutation} prompts an LLM to generate five plausible but flawed code variants per problem and retains only variants with non-trivial behavior (positive-test pass rate strictly between 0 and 100\%). Second, \emph{syntactic mutation} uses \CodeIn{cargo-mutants}~\cite{cargo_mutants} to generate additional code mutants; together, these two stages target up to 8$\times$ the positive set when enough distinct outputs are available. Finally, \emph{direct mutation} perturbs reference outputs without code, filling the remaining gap to the 10$\times$ target. Additional details are provided in Appendix~\ref{app:implementation_details}.

\textbf{Postcondition testing.} We use negative cases to validate postcondition completeness. Any postcondition that accepts an incorrect output is incomplete and must be revised. We focus on postconditions rather than preconditions because preconditions are simple input constraints that are easy to validate through manual review, whereas postconditions encode the full input--output relation and are hard to judge by inspection alone~\cite{richter2025beyond, vikram2023can, sun2024clover}. To scale this check, we implement \emph{Post2Exe}, which translates supported Verus postconditions into executable Rust programs and runs them on the negative tests. Post2Exe converts 85\% of benchmark postconditions; unsupported cases, such as those with unbounded quantifiers, are reviewed manually. This process identifies 60 incomplete postconditions, which are sent back to the agent for revision. Figure~\ref{fig:post_testing} shows one such case: the original specification for LeetCode 34 failed to require each returned index to be either $-1$ or non-negative, so mutated outputs such as $[-9,-8]$ were accepted. Adding the missing constraints restores completeness.

%% file: sections/4-benchmark.tex
\section{\benchmark}

\benchmark consists of \benchsize problems (722 from LeetCode and 284 from Codeforces). Each problem contains:
\begin{itemize}[topsep=0.2em, itemsep=0.1em, left=0pt]
    \item \emph{Natural language problem description}: the informal problem statement sourced from LeetCode or Codeforces. Each description is supplemented with starter code to support evaluation.
    \item \emph{Formal specification}: a ground-truth formal specification consisting of preconditions, which state the properties inputs must satisfy, and postconditions, which specify the desired relationship between inputs and outputs.
    \item \emph{Code}: ground-truth Rust code accepted by the online judge.
    \item \emph{Proof}: a ground-truth proof establishing that the code satisfies the specification.
    \item \emph{Positive and negative test cases}: positive test cases are valid input-output pairs, while negative test cases pair valid inputs with incorrect outputs for postcondition-completeness checks.
    \item \emph{Metadata}: problem ID, difficulty level, acceptance rate, and algorithm tags.
\end{itemize}

For lightweight evaluation, we also provide \emph{\benchmark Lite}, a 300-problem subset that closely follows the full benchmark's source, difficulty, and reference proof-size distributions. Appendix~\ref{app:vericontest_lite} describes its construction and reports model performance.

\subsection{Quality Assurance}

We enforce benchmark quality along five dimensions, with each benchmark instance reviewed by at least two human experts:

\begin{itemize}[topsep=0em, itemsep=0em, left=0pt]
    \item \emph{Code correctness and efficiency}: We submit all code to the online judge to ensure that it is accepted within the time and memory limits.
    \item \emph{Specification soundness}: Every problem includes a Verus proof certifying that the judge-accepted code satisfies the specification. Since the code is independently validated as correct, this establishes that the specification is sound with respect to the intended behavior.
    \item \emph{Specification completeness}: We verify that postconditions fully capture the intended requirements through both manual review and automated checking with negative test cases.
    \item \emph{Specification review}: We manually review each specification to avoid unnecessary implementation-specific constraints.
    \item \emph{High-quality test cases}: We include comprehensive positive and negative test cases for evaluating code correctness and specification completeness. 
\end{itemize}

\subsection{Benchmark Statistics \& Distribution}

\begin{figure*}[t]
\centering
\begin{minipage}[t]{0.52\textwidth}
\vspace{0pt}
\centering
\captionsetup{hypcap=false}
\captionof{table}{Statistics for \benchmark, including description length, lines of code/specification/proof, counts of loop invariants, assertions, and lemma functions, and generated tests.}
\label{tab:bench_statistics}
\small
\begin{tabular}{@{}lccc@{}}
\toprule
\textbf{Statistic} & \textbf{Median} & \textbf{Mean} & \textbf{Max} \\
\midrule
\# Words in Description & 206 & 237.7 & 902 \\
Lines of Code & 33 & 37.4 & 334 \\
Lines of Specification  & 26 & 31.6 & 220 \\
Lines of Proof & 94 & 157.9 & 1,769 \\
\midrule
\# Loop Invariants & 12 & 17.3 & 157 \\
\# Assertions & 17 & 32.0 & 359 \\
\# Lemma Functions & 1 & 3.5 & 60 \\
\midrule
\# Positive Tests & 268.5 & 246.4 & 395 \\
\# Negative Tests & 2,610 & 2,254.7 & 3,950 \\
\bottomrule
\end{tabular}
\end{minipage}\hfill
\begin{minipage}[t]{0.43\textwidth}
\vspace{0pt}
\centering
\includegraphics[width=0.86\linewidth]{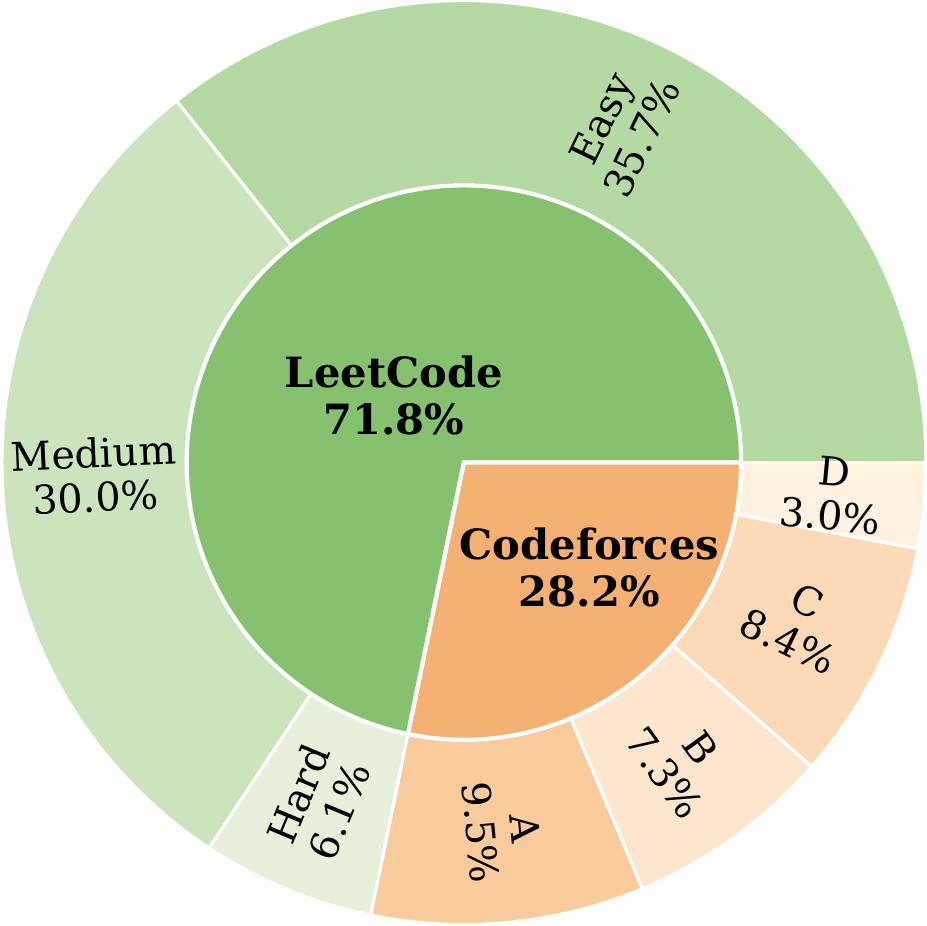}
\captionsetup{hypcap=false}
\captionof{figure}{Distribution of benchmark sources and difficulty levels.}
\label{fig:diff_dis}
\end{minipage}
\end{figure*}
    
\textbf{Benchmark statistics.} Table~\ref{tab:bench_statistics} summarizes the scale and complexity of \benchmark. Descriptions have a median length of 206 words; code, specifications, and proofs reach 334, 220, and 1,769 lines, respectively; and proofs require up to 157 loop invariants, 359 assertions, and 60 lemma functions. Each instance has a median of 268.5 positive and 2,610 negative tests. The negative-test count is not exactly 10$\times$ the positive-test count because Boolean or yes/no outputs leave only one or a few distinct incorrect outputs per input.

\textbf{Benchmark distribution.} Figure~\ref{fig:diff_dis} shows that \benchmark covers LeetCode Easy, Medium, and Hard tasks and Codeforces A--D tasks. We intentionally concentrate the distribution on problems that are challenging but still feasible for end-to-end verification. Including substantially harder problems would often require unsupported Rust features or data structures, such as priority queues and balanced trees, or proofs beyond the current capabilities of coding agents and human experts within a practical construction workflow.

%% file: sections/5-evaluation.tex
\section{Evaluation Setup}

\textbf{Models and prompting.} We evaluate ten state-of-the-art LLMs: GPT-5.5, GPT-5.4 mini, Claude Opus 4.7, Claude Sonnet 4.6, Gemini 3.1 Pro, Gemini 3 Flash, DeepSeek V4 Pro, DeepSeek V4 Flash, Qwen 3.6, and GLM-4.7-Flash. We use one-shot prompting with an example from Verus-Bench~\cite{yang2025autoverus} to standardize the output format without leaking benchmark content. Detailed model configurations, compute setup, costs, and temperature settings are provided in Appendix~\ref{app:evaluation_details}. Prompt templates are provided in Appendix~\ref{app:evaluation_prompts}.

\subsection{Evaluation Tasks}

\benchmark provides specifications, code, and proofs, enabling both isolated and compositional evaluation. We evaluate four core tasks: \textbf{SpecGen}, \textbf{CodeGen}, \textbf{ProofGen}, and \textbf{End2End}, with CodeGen decomposed into three input settings, yielding six subtasks:

\begin{itemize}[topsep=0.1em, itemsep=0.1em, left=0pt]
    \item \textbf{SpecGen} (\CodeIn{nl2spec}): given the natural language problem description, generate the specification.
    \item \textbf{CodeGen} consists of three subtasks that vary the input context:
    \begin{itemize}[topsep=0.1em, itemsep=0.1em, left=0pt]
        \item \CodeIn{nl2code}: given the natural language problem description, generate the executable Rust code.
        \item \CodeIn{spec2code}: given the ground-truth specification, generate executable Rust code satisfying it.
        \item \CodeIn{nl\_spec2code}: given both the natural language problem description and the ground-truth specification, generate the executable Rust code.
    \end{itemize}
    \item \textbf{ProofGen} (\CodeIn{spec\_code2proof}): given the ground-truth specification and executable code, generate the Verus proofs needed to verify the program.
    \item \textbf{End2End} (\CodeIn{end2end}): given only the natural language problem description, generate the full verified Verus program, including the specification, executable code, and proofs.
\end{itemize}

These settings isolate specification, code, and proof generation, compare CodeGen with different input contexts, and evaluate the full pipeline in \CodeIn{end2end}.

\subsection{Evaluation Metrics}
\label{sec:evaluation_metrics}

\textbf{SpecGen.} We check preconditions and postconditions separately. A generated precondition must be neither weaker nor stronger than the ground truth. Let $P$ and $\widehat{P}$ denote the ground-truth and generated preconditions over inputs $\overline{x}$. We check both directions:
\[
    \forall \overline{x}.~ P(\overline{x}) \Rightarrow \widehat{P}(\overline{x})
    \qquad\mbox{and}\qquad
    \forall \overline{x}.~ \widehat{P}(\overline{x}) \Rightarrow P(\overline{x}).
\]
The first implication rules out generated preconditions that reject valid inputs (completeness), while the second rules out those that admit inputs excluded by the ground truth (soundness). We ask an LLM to generate proofs and use Verus to check both obligations. Details are provided in Appendix~\ref{app:specgen_metric}.

Preconditions usually encode simple input constraints, making their equivalence easier to verify. Postconditions express the full input--output relation, whose equivalence cannot be reliably verified by current LLMs. We therefore use Post2Exe to convert the generated postcondition into an executable Rust program, which we run on positive tests for soundness and negative tests for completeness. If precondition verification or Post2Exe conversion fails, human experts manually compare the generated specification with the ground truth.

\textbf{CodeGen.} We run generated code on positive tests with a one-second timeout. Code is correct only if all tests pass within the timeout. 

\textbf{ProofGen.} With the ground-truth specification and executable code fixed, the model should only add proofs. We reject outputs that change the specification or code, and run Verus with \CodeIn{-{}-no-cheating} to determine whether the proofs are accepted.

\textbf{End-to-End.} A generated program is end-to-end correct only if its specification passes SpecGen evaluation, its executable code passes all positive tests within the per-case timeout, and Verus accepts its proofs against the generated specification with \CodeIn{-{}-no-cheating}.

We report pass@1~\cite{chen2021evaluating} for SpecGen, CodeGen, ProofGen, and End2End. For ProofGen, we also report pass@k~\cite{chen2021evaluating} and repair@k for $k=1,\ldots,10$ on a stratified subset of 200 problems. The repair@k metric measures whether a proof can be accepted after up to $k$ rounds of repair. In each round, the model receives the specification, code, previous failed proof, and Verus error messages, then produces a revised proof.

\section{Evaluation Results}

\input{tables/main_results.tex}

Table~\ref{tab:main_results} reports pass@1 on the six subtask settings, grouped into SpecGen (\CodeIn{nl2spec}), CodeGen (\CodeIn{nl2code}, \CodeIn{spec2code}, \CodeIn{nl\_spec2code}), ProofGen (\CodeIn{spec\_code2proof}), and End2End (\CodeIn{end2end}). Figure~\ref{fig:proofgen_passk_repairk} further examines ProofGen with independent sampling and iterative proof repair. Because pass@k and repair@k require repeated model calls and incur high API cost, we report them only for six models on a stratified 200-problem subset that preserves the benchmark's source and difficulty distribution. We organize the findings along these four tasks.

\subsection{SpecGen}

\noindent\emph{\textbf{F1: SpecGen is challenging for current LLMs.}}
Across models, \CodeIn{nl2spec} is much harder than standard code generation. GPT-5.5 achieves 57.16\% on \CodeIn{nl2spec}, compared with 97.51\% on \CodeIn{nl2code}; Claude Opus 4.7 drops from 95.43\% to 27.24\%, and Gemini 3.1 Pro drops from 94.14\% to 22.56\%. Open-source models follow the same pattern at lower absolute accuracy: Qwen 3.6 reaches 82.50\% on \CodeIn{nl2code} but only 0.40\% on \CodeIn{nl2spec}. These results indicate that current LLMs remain much less familiar with formal specifications than with code generation.

\noindent\emph{\textbf{F2: CodeGen ability helps SpecGen, but does not guarantee it.}}
Strong \CodeIn{nl2code} models often lead \CodeIn{nl2spec}: GPT-5.5 ranks first on both tasks (97.51\% and 57.16\%), and Claude Opus 4.7 ranks second on both (95.43\% and 27.24\%). However, this relationship is not reliable across all models. GPT-5.4 mini achieves 83.90\% on \CodeIn{nl2code} but only 7.85\% on \CodeIn{nl2spec}, and Qwen 3.6 similarly splits between 82.50\% and 0.40\%. Conversely, Gemini 3 Flash is much weaker on \CodeIn{nl2code} than DeepSeek V4 Pro (53.08\% vs. 92.05\%) but higher on \CodeIn{nl2spec} (22.66\% vs. 12.13\%). Thus, SpecGen benefits from CodeGen but also requires the ability to express input--output behavior as formal specifications.

\subsection{CodeGen}

\noindent\emph{\textbf{F1: Models perform strongly on CodeGen.}}
CodeGen is the strongest task. On \CodeIn{nl2code}, GPT-5.5 reaches 97.51\%, followed by Claude Opus 4.7 at 95.43\%, Gemini 3.1 Pro at 94.14\%, and Claude Sonnet 4.6 at 93.84\%. Open-source models remain competitive: DeepSeek V4 Pro reaches 92.05\%, while Qwen 3.6 reaches 82.50\% with only 27B parameters. Overall, eight of the ten models exceed 80\%, showing that strong performance extends across multiple model families. Generating executable Rust code from natural language is therefore relatively mature compared with specification and proof generation.

\noindent\emph{\textbf{F2: Specifications remain difficult for models to use as CodeGen input.}}
Replacing natural language with a formal specification reduces CodeGen accuracy for all models. GPT-5.5 falls from 97.51\% on \CodeIn{nl2code} to 77.44\% on \CodeIn{spec2code}, Claude Opus 4.7 falls from 95.43\% to 77.24\%, and Qwen 3.6 drops from 82.50\% to 43.94\%. This suggests that current LLMs generate code more effectively from natural language than from Verus specifications alone. This pattern is consistent with a training-distribution mismatch: current LLMs are exposed to abundant natural-language programming problems and solutions, but far fewer specification--code pairs. Consequently, models can exploit familiar problem statements in \CodeIn{nl2code}, but struggle when the specification is the only input and the required behavior must be inferred from formal logical constraints.

\noindent\emph{\textbf{F3: Natural language helps specification-based CodeGen, but specifications hurt natural-language CodeGen.}}
The three CodeGen settings follow a consistent ordering: \CodeIn{spec2code} $<$ \CodeIn{nl\_spec2code} $<$ \CodeIn{nl2code}. Adding natural language to a specification improves performance over \CodeIn{spec2code}: GPT-5.5 rises from 77.44\% to 84.49\%, and Gemini 3.1 Pro rises from 80.62\% to 89.86\%. However, adding a specification to the natural language description reduces accuracy compared with \CodeIn{nl2code}: Claude Opus 4.7 drops from 95.43\% on \CodeIn{nl2code} to 80.62\% on \CodeIn{nl\_spec2code}. This again suggests a training-distribution mismatch: natural language provides familiar semantic and algorithmic cues, while formal specifications introduce unfamiliar syntax and type-level distinctions. A common failure is copying Verus-only ghost types such as \CodeIn{int} and \CodeIn{nat} into executable code, causing syntax errors even when the algorithm is correct. We provide detailed error analysis in Appendix~\ref{app:error_analysis}.

\subsection{ProofGen}

\begin{figure*}[t]
    \centering
    \includegraphics[width=\textwidth]{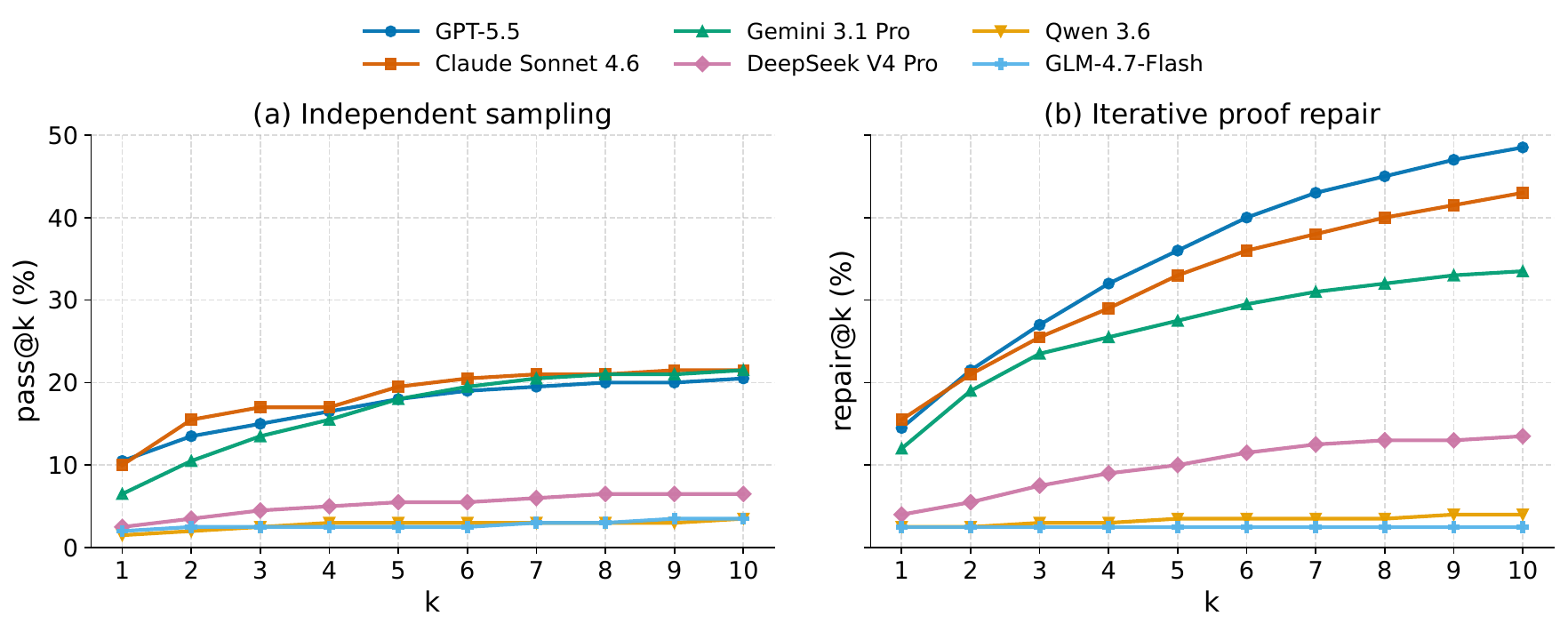}
    \caption{ProofGen performance under independent sampling and iterative proof repair on the stratified 200-problem subset.}
    \label{fig:proofgen_passk_repairk}
\end{figure*}

\noindent\emph{\textbf{F1: ProofGen is the major bottleneck for verifiable code generation.}}
Even with the ground-truth specification and executable code, all models have low accuracy on \CodeIn{spec\_code2proof}. GPT-5.5 reaches 13.82\%, Claude Sonnet 4.6 reaches 13.52\%, and most models remain below 8\%, including GPT-5.4 mini (6.36\%), DeepSeek V4 Pro (5.47\%), and Qwen 3.6 (4.57\%). Unlike CodeGen or SpecGen, this setting does not require models to discover an algorithm or formalize the problem requirements. Instead, models must supply the loop invariants, assertions, and lemma functions needed to establish that the provided code satisfies its specification. The low success rates indicate that access to a correct implementation and its specification is insufficient for current models to reliably construct an accepted proof. This gap highlights the distinction between generating executable code and making its correctness explicit to a verifier, making proof generation a central bottleneck for verifiable code generation.

\noindent\emph{\textbf{F2: Independent sampling and iterative proof repair both help ProofGen.}}
Figure~\ref{fig:proofgen_passk_repairk} shows that both independent sampling and iterative repair improve ProofGen as the attempt budget increases for five of the six models. For GPT-5.5, pass@k rises from 10.5\% at $k=1$ to 20.5\% at $k=10$, while repair@k rises from 14.5\% to 48.5\%. Claude Sonnet 4.6, Gemini 3.1 Pro, DeepSeek V4 Pro, and Qwen 3.6 follow the same trend. GLM-4.7-Flash is the exception: pass@k increases from 2.0\% to 3.5\%, while repair@k remains at 2.5\%. Thus, some failed proofs are recoverable, but most remain unresolved after ten attempts or repair rounds.

\noindent\emph{\textbf{F3: Proof repair is more effective than independent sampling.}}
Repair outperforms independent sampling at every reported $k$ for GPT-5.5, Claude Sonnet 4.6, Gemini 3.1 Pro, and DeepSeek V4 Pro, and the gap widens with $k$. For GPT-5.5, repair gains 34.0 percentage points from $k=1$ to $k=10$, compared with 10.0 points for sampling. Its gain decreases from 7.0 points between $k=1$ and $k=2$ to 1.5 points between $k=9$ and $k=10$. Repair therefore narrows the ProofGen gap without closing it: GPT-5.5 reaches 48.5\% at repair@10, still below its 97.51\% CodeGen pass@1 accuracy. On the same subset, Qwen 3.6 reaches 4.0\% at repair@10 versus 3.5\% at pass@10, whereas GLM-4.7-Flash reaches 2.5\% versus 3.5\%.

\subsection{End-to-End}

\noindent\emph{\textbf{F1: End-to-end verifiable code generation is far from solved.}}
All models have very low accuracy on \CodeIn{end2end}. GPT-5.5 reaches only 6.76\%, followed by Gemini 3.1 Pro at 3.68\%, Claude Opus 4.7 at 3.58\%, and Claude Sonnet 4.6 at 3.48\%. The remaining six models score below 1.5\%, with Qwen 3.6 and GLM-4.7-Flash reaching only 0.20\% and 0.70\%, respectively. This sharp drop from isolated subtasks shows that current LLMs cannot reliably compose specification, code, and proof generation into a complete verified program.

\noindent\emph{\textbf{F2: End-to-end success requires all three capabilities to align.}}
End-to-end success requires the specification, code, and proof to all pass their checks for the same problem. GPT-5.5 leads \CodeIn{nl2spec} (57.16\%), \CodeIn{nl2code} (97.51\%), and \CodeIn{spec\_code2proof} (13.82\%), yet reaches only 6.76\% on \CodeIn{end2end}. Claude Opus 4.7 shows the same compounding effect: despite strong CodeGen (95.43\%), lower SpecGen (27.24\%) and ProofGen (12.52\%) accuracies limit end-to-end success to 3.58\%. Thus, any failure in specification, code, or proof generation can bottleneck the full pipeline. 

\textbf{Additional analysis.} Appendix~\ref{app:additional_evaluation_analysis} presents difficulty-stratified results, detailed error analysis, a case study, and statistical uncertainty and paired comparisons. Limitations and data contamination are discussed in Appendix~\ref{app:limitations}.

%% file: tables/main_results.tex
\begin{table*}[t]
\centering
\caption{Main results of pass@1 accuracy across six evaluation tasks. The \textbf{\highlight{Green!15}{highest}} and \highlight{Cyan!15}{second-highest} accuracies are highlighted.}
\label{tab:main_results}
\small
\scalebox{0.88}{
\begin{tabular}{l|c|ccc|c|c}
\toprule
\textbf{Model} &
\multicolumn{1}{c}{\textbf{SpecGen}} &
\multicolumn{3}{c}{\textbf{CodeGen}} &
\multicolumn{1}{c}{\textbf{ProofGen}} &
\multicolumn{1}{c}{\textbf{End-to-End}} \\
\cmidrule(lr){2-2}
\cmidrule(lr){3-5}
\cmidrule(lr){6-6}
\cmidrule(lr){7-7}
& nl2spec
& nl2code
& spec2code
& nl\_spec2code
& spec\_code2proof
& end2end \\
\midrule
GPT-5.5 & \textbf{\highlight{Green!15}{57.16}} & \textbf{\highlight{Green!15}{97.51}} & 77.44 & \highlight{Cyan!15}{84.49} & \textbf{\highlight{Green!15}{13.82}} & \textbf{\highlight{Green!15}{6.76}} \\
GPT-5.4 mini & 7.85 & 83.90 & 54.57 & 63.02 & 6.36 & 1.49 \\
\midrule
Claude Opus 4.7 & \highlight{Cyan!15}{27.24} & \highlight{Cyan!15}{95.43} & 77.24 & 80.62 & 12.52 & 3.58 \\
Claude Sonnet 4.6 & 17.30 & 93.84 & \textbf{\highlight{Green!15}{82.01}} & 83.80 & \highlight{Cyan!15}{13.52} & 3.48 \\
\midrule
Gemini 3.1 Pro & 22.56 & 94.14 & \highlight{Cyan!15}{80.62} & \textbf{\highlight{Green!15}{89.86}} & 12.92 & \highlight{Cyan!15}{3.68} \\
Gemini 3 Flash & 22.66 & 53.08 & 32.50 & 36.58 & 6.16 & 1.29 \\
\midrule
DeepSeek V4 Pro & 12.13 & 92.05 & 66.80 & 73.06 & 5.47 & 1.29 \\
DeepSeek V4 Flash & 9.74 & 82.21 & 64.31 & 64.51 & 6.36 & 1.19 \\
\midrule
Qwen 3.6 & 0.40 & 82.50 & 43.94 & 51.59 & 4.57 & 0.20 \\
\midrule
GLM-4.7-Flash & 3.68 & 43.34 & 17.10 & 22.17 & 5.27 & 0.70 \\
\bottomrule
\end{tabular}
}
\vspace{-2mm}
\end{table*}

%% file: sections/6-conclusion.tex
\section{Conclusion}

We introduced \benchmark, a benchmark of \benchsize competitive-programming problems from LeetCode and Codeforces for evaluating verifiable code generation in Rust with Verus. \benchmark supports evaluating specification, code, and proof generation both individually and together in end-to-end verified program synthesis. Our evaluation of ten state-of-the-art LLMs reveals a sharp gap between ordinary coding and verifiable code generation: the strongest model reaches 97.51\% on natural-language-to-code generation, but only 57.16\% on specification generation, 13.82\% on proof generation, and 6.76\% end-to-end. These results show that current models struggle to generate formal specifications and proofs, which remain major bottlenecks for reliable end-to-end verifiable code generation. We hope that \benchmark will serve as both a rigorous framework for evaluating verifiable code generation and a valuable source of training data for improving future models.

%% file: sections/appendix.tex
\begingroup
\raggedbottom
\setlength{\textfloatsep}{10pt plus 2pt minus 2pt}
\setlength{\floatsep}{8pt plus 2pt minus 2pt}
\setlength{\intextsep}{10pt plus 2pt minus 2pt}
\renewcommand{\topfraction}{0.9}
\renewcommand{\bottomfraction}{0.8}
\renewcommand{\textfraction}{0.08}
\renewcommand{\floatpagefraction}{0.8}

\section{Benchmark Construction}
\label{app:benchmark_details}

\subsection{Construction Filtering and Attrition}
\label{app:cr_attrition}

Table~\ref{tab:cr_attrition} reports the number of problems retained at each construction stage. The LeetCode pool contains 3,862 problems as of March 13; after excluding premium problems, we screen all 3,106 remaining candidates. For Codeforces, the rating threshold of 1,900 leaves 6,062 candidates, of which we screen 548 within the construction budget. The Codeforces counts therefore describe this screened subset rather than the full candidate pool.

\input{tables/construction_attrition}

Across both sources, screening excludes 2,294 problems, and 354 additional problems fail verification within the construction budget. Table~\ref{tab:cr_exclusions} groups these cases by reason. Unsupported features and complex data structures account for most screening exclusions, while greedy and dynamic-programming optimality proofs are the largest source of verification failures. Thus, the construction pipeline favors problems that fit the Rust/Verus subset and admit proofs within the available budget.

\input{tables/construction_exclusions}

Table~\ref{tab:cr_filter_distribution} compares the recorded distributions before and after filtering. Filtering shifts both sources toward easier problems and algorithmic categories supported by the construction pipeline. These percentages describe the filtering stage; the final benchmark is further affected by verification failures, and its stratum sizes are reported in Appendix~\ref{app:cr_difficulty}.

\input{tables/filter_distribution}

\subsection{Expert Review and Specification Validation}
\label{app:cr_expert_review}

\textbf{Reviewers and rubric.} Phase~I involves four experts, each with at least five years of computer science experience and a strong background in Verus and formal methods. They write and cross-check the seed specifications, code, and proofs. Phase~II involves six experts, with each problem independently reviewed by at least two experts. The reviewers check code correctness and efficiency through online-judge acceptance, compare the preconditions and postconditions with the intended problem, and inspect the Verus proof. Each reviewer records an accept or revise verdict with a written reason.

\textbf{Resolution and effort.} A third reviewer compares the two verdicts. Two accept verdicts lead to acceptance, while two revise verdicts trigger correction. When the verdicts disagree, the reviewers discuss the problem and reach a joint verdict. Every revised problem is reviewed again. Phase~II expert review takes approximately seven minutes per problem on average, in addition to the two 20-minute construction stages.

\textbf{Corrections.} The validation process identifies 132 specifications that require correction, corresponding to 13.1\% of the 1,006-problem benchmark. Of these, 45 are found during Phase~II human artifact validation, 60 through Post2Exe postcondition testing, 13 through manual review of non-convertible postconditions, and 14 through two-expert precondition review.

\textbf{Reviewer agreement.} On the 152 postconditions that Post2Exe cannot convert, the two reviewers initially agree on 131 cases (86.2\%). They resolve the remaining 21 cases through discussion, identifying 10 incomplete postconditions. For preconditions, the reviewers agree on 993 of 1,006 cases (98.7\%); all 13 disputed cases require correction.

\textbf{Example correction.} In LeetCode 1276, the original specification constrains the tomato and cheese counts only when the result contains two entries. It consequently permits an empty result even when a feasible solution exists. The reviewers initially disagree on this case, then replace the one-way implication with a biconditional (Figure~\ref{fig:cr_spec_correction}).
\begin{figure}[t]
\centering
\begin{tcolorbox}[colback=white, colframe=black, boxrule=0.6pt, left=0pt, right=0pt, top=0pt, bottom=0pt, arc=0pt, outer arc=0pt]
\begin{lstlisting}[style=verusblock, firstnumber=1]
// Original: a two-entry result implies feasibility.
result@.len() == 2 ==> (
    tomato_slices as int % 2 == 0
    && 2 * (cheese_slices as int) <= tomato_slices as int
    && tomato_slices as int <= 4 * (cheese_slices as int))

// Revised: feasibility also requires a two-entry result.
result@.len() == 2 <==> (
    tomato_slices as int % 2 == 0
    && 2 * (cheese_slices as int) <= tomato_slices as int
    && tomato_slices as int <= 4 * (cheese_slices as int))
\end{lstlisting}
\end{tcolorbox}
\caption{Original and revised specification clauses for LeetCode 1276.}
\label{fig:cr_spec_correction}
\end{figure}
This change rejects an empty result whenever a solution exists; the remaining clauses constrain the returned counts.

\textbf{Post2Exe coverage.} Post2Exe checks 854 of 1,006 postconditions (84.9\%) and detects 60 incomplete cases (7.0\%). The remaining 152 postconditions use unsupported constructs, such as unbounded quantifiers, quantifiers over sequence-valued variables, or specification-helper stubs, and are reviewed manually. This review identifies 13 additional incomplete postconditions (8.6\%).

Manual review is more common for harder problems. For LeetCode, the non-conversion rate increases from 5.3\% on Easy problems to 12.3\% on Medium and 18.0\% on Hard problems. Hard problems constitute 8.4\% of the LeetCode benchmark but 16.4\% of its non-convertible subset. For Codeforces, the rate is 27\% for problems rated at most 1,300 and 41\% for problems rated at least 1,400. Harder problems tend to use more expressive postconditions that Post2Exe does not yet support.

The two validation routes detect incomplete postconditions at similar rates (7.0\% and 8.6\%), although they cover different difficulty distributions and use different checking procedures. Together, online-judge acceptance, Verus verification, negative tests, and manual review check complementary aspects of each benchmark problem.

\subsection{Test Case Generation Details}
\label{app:implementation_details}

\subsubsection{Positive Test Case Generation}

\textbf{Random test case generation.} For each problem, we synthesize a Verus-verified test-input generator from the problem description, the Verus specification, and the plain Rust function signature. The generator takes construction parameters and returns an input for the target program. Its postcondition implies the target program's precondition, so Verus verification ensures that every generated input is valid. The model writes both the generator and the proof of its correctness.

Many generators include multiple verified synthesis modes, selected through an additional control parameter, to produce structurally distinct valid instances. For example, array generators may insert, remove, swap, or perturb elements while preserving invariants such as bounds or sortedness. After verification, we compile and run the generator with randomly sampled parameters, compute the corresponding outputs with the accepted reference implementation, and deduplicate the resulting input-output pairs as random positive test cases.

\textbf{Adversarial test case generation.} We generate adversarial test cases with the same Verus-verified generator framework, but change the prompt objective from broad coverage to failure-mode targeting. Specifically, we ask the model to analyze the full specification and problem description: the \CodeIn{requires} clause determines the valid input space, while the \CodeIn{ensures} clause and task semantics guide the failure modes to target. These failure modes include off-by-one errors, mishandled duplicates, boundary-condition mistakes, overflow-sensitive computations, and incorrect handling of sortedness or edge cases. As in random generation, the generator's postcondition implies the target program's precondition, and Verus verification ensures that every produced input is valid.

The adversarial cases are generated by dedicated modes that target the identified failure modes, such as minimal and maximal sizes, boundary values, degenerate structures, duplicate-heavy inputs, and inputs that stress problem-specific invariants. For constraints that require the existence of a witness, the generator is instructed to construct such witnesses explicitly. After verification, we compile and run the generator with a fixed seed, compute the corresponding outputs with the accepted reference implementation, and deduplicate the resulting input-output pairs as adversarial positive test cases.

\subsubsection{Negative Test Case Generation}

This section provides the implementation details of the negative test case pipeline introduced in the Benchmark Construction section (Phase~III). For every problem, we target a negative set of size $10\times$ the positive set. We first collect incorrect outputs from semantic and syntactic code mutants, targeting up to $8\times$ the positive set when enough distinct outputs are available, and then use direct output mutation to fill the remaining gap to the $10\times$ target.

\textbf{Semantic mutation.} For each problem, we prompt an LLM with the natural-language problem description and the verified Rust solution, and ask for five subtly broken variants that preserve the original function and helper signatures but inject a partial bug. The prompt enumerates a fixed catalogue of plausible engineering mistakes (off-by-one, wrong comparator, swapped indices, missing edge case, wrong initialization, accumulator reset, etc.) and requires the five variants to draw on \emph{distinct} categories, which discourages near-duplicates and broadens the bug distribution. A variant is retained only when its pass rate on the positive set lies strictly between $0\%$ and $100\%$. From every retained variant, we record the inputs on which its output disagrees with the reference, together with the resulting incorrect outputs, and contribute these pairs to the negative set.

\textbf{Syntactic mutation.} After semantic mutation, we use \CodeIn{cargo-mutants}~\cite{cargo_mutants} to move the negative set toward the $8\times$ target. It enumerates local AST edits, including operator swaps, negated conditionals, and return-value replacements. For each retained mutant, we record inputs whose outputs differ from the reference and add these incorrect outputs as additional negative cases, using the same per-input deduplication.

\textbf{Direct mutation.} This final stage perturbs the reference output $y$ of each positive case $(x, y)$ without invoking any code, filling the remaining gap to the $10\times$ target. We dispatch on the runtime type of $y$ and route each case to a type-specific generator. Perturbations include:
\begin{itemize}[nosep, left=0pt]
    \item \emph{Numerical} ($v \in \mathbb{Z}, \mathbb{R}$): additive deltas from $\{\pm 1, \pm 2, \pm 5, \pm 10, \pm 100\}$, sign flip, zero, doubling, halving, and bounded random offsets.
    \item \emph{Boolean}: logical negation.
    \item \emph{String}: empty string, prefix/suffix character insertion, head/tail truncation, reversal, single-position character replacement (a-z), and adjacent-character swap.
    \item \emph{Vector of int}: per-element $\pm 1$, pairwise swap, prepend/append, reversal, ascending and descending sort, replace one element with $0$, global $\pm 1$ shift, and element-wise negation.
    \item \emph{Vector of bool / string / matrix of int}: analogous variants (per-element negation, single-element string mutation, single-cell delta on 2-D matrices), with a generic fallback (drop first/last, reverse, empty) for unrecognized list element types.
\end{itemize}
We balance the negative set by drawing one mutation per positive input in round-robin order rather than exhausting all mutations of a single anchor case, and deduplicate against both the reference output and every mutation already retained for the same input.

\section{Evaluation Details}
\label{app:evaluation_details}

\subsection{Experimental Setup}

\textbf{Machine.} All experiments are run on a machine with Ubuntu 22.04.5 LTS, 192-core CPUs, and 512GB RAM. For open-source models run locally, we use NVIDIA RTX PRO 6000 Blackwell GPUs with 96 GB VRAM.

\textbf{Model configurations.} Table~\ref{tab:llms} reports each evaluated model's vendor, checkpoint, access mode, input/output price, and total API cost when applicable.

\input{tables/llms.tex}

\textbf{Temperature.} For models that expose a temperature parameter, we set temperature to 0 for pass@1 to make the evaluation deterministic, and to 0.7 for pass@k and repair@k to allow sampling diversity. For models that do not expose user-configurable temperature settings (Claude Opus 4.7 and GPT-5.5), we use the default setting.

\subsection{Precondition Evaluation Details for Specification Generation}
\label{app:specgen_metric}

Section~\ref{sec:evaluation_metrics} defines the two implications required for precondition equivalence. For each generated precondition, we use GPT-5.5 to generate Verus proofs for these obligations. We allow up to five proof-repair rounds, meaning that the model can attempt to repair a proof up to five times if it fails to verify. We run Verus on the resulting proofs; if both obligations verify, the generated precondition is accepted as equivalent to the ground truth. If verification fails, human experts manually compare the generated precondition with the ground truth.

\section{Additional Evaluation Analysis}
\label{app:additional_evaluation_analysis}

\subsection{Difficulty-Stratified Performance}
\label{app:cr_difficulty}

Table~\ref{tab:cr_difficulty} reports pass@1 for GPT-5.5 and Claude Sonnet 4.6 across six difficulty strata. LeetCode contributes 359 Easy, 302 Medium, and 61 Hard problems; Codeforces contributes 158 problems rated 800--1000, 72 rated 1100--1300, and 54 rated 1400--1900.

\input{tables/difficulty_stratified}

\noindent\emph{\textbf{F1: CodeGen remains strong across difficulty levels.}}
GPT-5.5 achieves at least 95.8\% on every LeetCode stratum and at least 98.1\% on every Codeforces stratum. Claude Sonnet 4.6 also remains above 75\% across all six strata. These scores are consistent with the strong aggregate CodeGen results in Table~\ref{tab:main_results}.

\noindent\emph{\textbf{F2: ProofGen shows the clearest decline with difficulty.}}
For GPT-5.5, ProofGen falls from 16.7\% on LeetCode Easy to 6.6\% on Hard, and from 30.4\% on Codeforces 800--1000 to 3.7\% on 1400--1900. Claude Sonnet 4.6 follows the same trend, falling from 15.0\% to 4.9\% on LeetCode and from 25.9\% to 1.9\% on Codeforces. SpecGen and End2End do not decrease monotonically in every stratum, but End2End reaches 1.6\% on LeetCode Hard for both models. These results show that proof generation becomes particularly challenging for harder problems, which current models still struggle to solve effectively.

\subsection{Error Analysis}
\label{app:error_analysis}

\begin{figure*}[t]
    \centering
    \includegraphics[width=\textwidth]{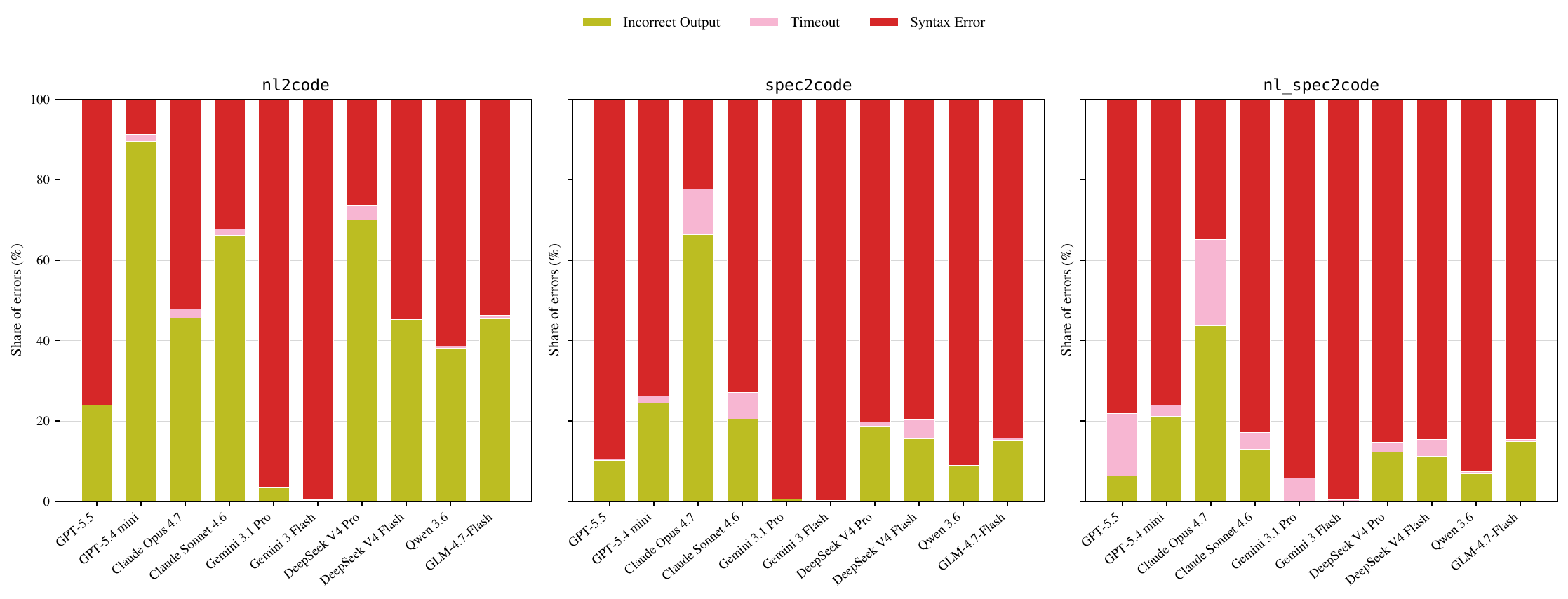}
    \caption{Error composition for the three CodeGen subtasks (\CodeIn{nl2code}, \CodeIn{spec2code}, \CodeIn{nl\_spec2code}) for the ten evaluated models. Each bar normalizes over failed attempts only and shows the share of errors in each category: \emph{Incorrect Output}, \emph{Timeout}, and \emph{Syntax Error}. The corresponding pass rates are reported in Table~\ref{tab:main_results}.}
    \label{fig:error_analysis_codegen}
\end{figure*}

\begin{figure*}[t]
    \centering
    \includegraphics[width=\textwidth]{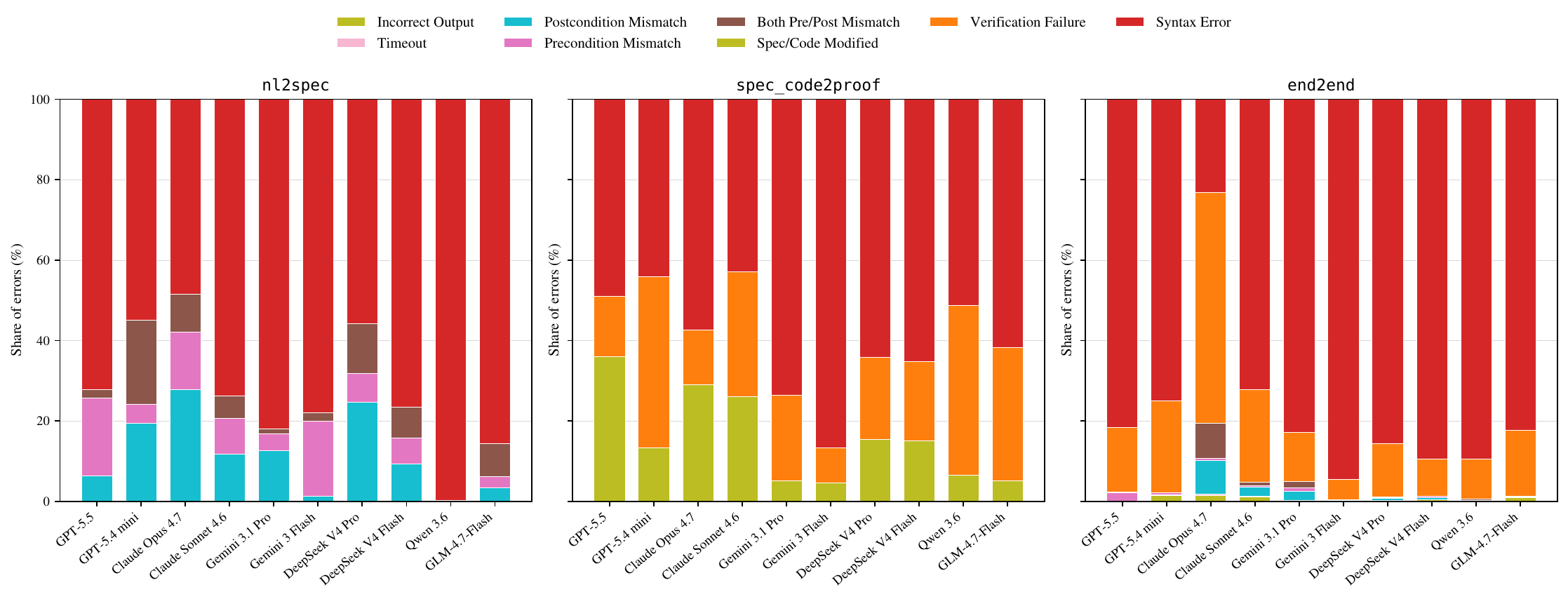}
    \caption{Error composition for SpecGen (\CodeIn{nl2spec}), ProofGen (\CodeIn{spec\_code2proof}), and the End2End pipeline (\CodeIn{end2end}) for the ten evaluated models. Each bar normalizes over failed attempts only. \emph{Precondition Mismatch}, \emph{Postcondition Mismatch}, and \emph{Both Pre/Post Mismatch} apply to \CodeIn{nl2spec} and \CodeIn{end2end}; \emph{Spec/Code Modified} applies only to \CodeIn{spec\_code2proof}; \emph{Incorrect Output} and \emph{Timeout} apply only to \CodeIn{end2end} in this figure. The corresponding pass rates are reported in Table~\ref{tab:main_results}.}
    \label{fig:error_analysis_specproofe2e}
\end{figure*}

To better understand the pass rates in Table~\ref{tab:main_results}, we classify every failed attempt on each of the six subtasks into a set of error categories. Figures~\ref{fig:error_analysis_codegen} and~\ref{fig:error_analysis_specproofe2e} report the resulting distributions. The error categories are:

\begin{itemize}[left=0pt]
    \item \textbf{Incorrect Output}: the program compiles, but at least one test case returns a wrong output.
    \item \textbf{Timeout}: execution exceeds the per-case time limit, indicating that the generated implementation is too slow for the test suite.
    \item \textbf{Verification Failure}: the generated program is syntactically valid, but Verus fails to verify it due to precondition, postcondition, loop invariant, assertion, arithmetic-overflow, unsupported-feature, or termination-witness failures.
    \item \textbf{Syntax Error}: the output contains Rust or Verus syntax errors, such as parser failures, rustc errors, or missing trigger annotations.
    \item \textbf{Precondition Mismatch}, \textbf{Postcondition Mismatch}, and \textbf{Both Pre/Post Mismatch}: the program parses, but its precondition fails the precondition equivalence check (Appendix~\ref{app:specgen_metric}), its postcondition fails the positive/negative test check, or both.
    \item \textbf{Spec/Code Modified} (\CodeIn{spec\_code2proof} only): specification or executable code is modified during proof generation.   
\end{itemize}

We organize the findings along five observations.

\noindent\emph{\textbf{F1: The dominant failure mode shifts once Verus is involved.}}
On \CodeIn{nl2code}, where the model emits plain Rust, errors are dominated by \emph{Incorrect Output} for several models: 89.5\% of GPT-5.4 mini's errors, 70.0\% of DeepSeek V4 Pro's, and 66.1\% of Claude Sonnet 4.6's are wrong outputs. Once the input switches to a Verus specification, \emph{Syntax Error} becomes the leading error category on \CodeIn{spec2code} for nearly every model: 99.7\% of Gemini 3 Flash's errors, 91.1\% of Qwen 3.6's, and 84.2\% of GLM-4.7-Flash's errors are syntactic. This leaves only a small share of failures that reach execution and expose an algorithmic bug. On \CodeIn{spec\_code2proof}, \emph{Syntax Error} remains the largest category for every model, but a different error pattern appears for the strongest models: \emph{Spec/Code Modified} accounts for 36.0\% of GPT-5.5's errors, 29.0\% of Claude Opus 4.7's, and 26.0\% of Claude Sonnet 4.6's, indicating that these models often modify the provided obligation rather than only adding the required proof. On \CodeIn{end2end}, where specification, code, and proof must all be produced together, \emph{Verification Failure} reaches 57.5\% of Claude Opus 4.7's errors, while the other models remain dominated by \emph{Syntax Error}. This suggests that Claude Opus 4.7 has a stronger command of Verus syntax and consistently produces parsable programs.

\noindent\emph{\textbf{F2: Frontier and open-source models fail at different stages.}}
Frontier models more often reach semantic or verification failures, while smaller open-source models are more frequently blocked by \emph{Syntax Error}. This split appears once Verus is required: on \CodeIn{end2end}, the share of \emph{Syntax Error} reaches 94.6\% for Gemini 3 Flash, 89.5\% for Qwen 3.6, and 82.4\% for GLM-4.7-Flash, while Claude Opus 4.7 keeps this share at 23.1\% and has most of its errors in \emph{Verification Failure}. This split indicates that frontier models are more often bottlenecked by specification and proof reasoning, whereas open-source models are more often bottlenecked by Verus syntax.

\noindent\emph{\textbf{F3: Timeouts emerge when a specification is added to the CodeGen input.}}
On \CodeIn{nl2code}, \emph{Timeout} is negligible. On \CodeIn{spec2code} and \CodeIn{nl\_spec2code}, however, the share of \emph{Timeout} grows for several of the strongest models, most sharply on \CodeIn{nl\_spec2code}. Claude Opus 4.7 produces 42 timeout cases, and GPT-5.5 produces 24, out of 1,006 \CodeIn{nl\_spec2code} problems. As shares of errors, \emph{Timeout} reaches 21.5\% for Claude Opus 4.7 and 15.4\% for GPT-5.5; on the same task, \emph{Incorrect Output} and \emph{Timeout} together account for 65.1\% of Claude Opus 4.7's errors (43.6\% \emph{Incorrect Output} and 21.5\% \emph{Timeout}). This pattern suggests that models struggle to bridge Verus specifications and efficient implementations: specifications often state behavior with quantifiers (\CodeIn{forall}, \CodeIn{exists}), and models will translate these logical conditions directly into executable code instead of synthesizing an efficient algorithm.

\noindent\emph{\textbf{F4: SpecGen failures are dominated by syntactic errors.}}
On \CodeIn{nl2spec}, \emph{Syntax Error} accounts for 72.3\% of GPT-5.5's errors, 55.0\% of GPT-5.4 mini's, 48.5\% of Claude Opus 4.7's, and 99.8\% of Qwen 3.6's. Among parsable specifications, which component fails varies across frontier models: Claude Opus 4.7 produces 27.8\% \emph{Postcondition Mismatch} and 14.3\% \emph{Precondition Mismatch}, while GPT-5.5 produces 6.4\% and 19.3\%. Weaker models more often fail on both components at once; for example, GPT-5.4 mini assigns 21.0\% of its errors to \emph{Both Pre/Post Mismatch}. Improving \CodeIn{nl2spec} therefore requires both better Verus syntax understanding and better grounding of formal specifications in the intended input--output behavior.

\noindent\emph{\textbf{F5: Verification failure exposes the core difficulty in ProofGen and End2End.}}
\emph{Verification Failure} becomes a dominant signal on \CodeIn{spec\_code2proof} for models that usually preserve the provided specification and executable code: it accounts for 42.1\% of Qwen 3.6's errors and 42.5\% of GPT-5.4 mini's, while \emph{Spec/Code Modified} stays below 25\% for both. On \CodeIn{end2end}, \emph{Verification Failure} accounts for 57.5\% of Claude Opus 4.7's errors, the largest \emph{Verification Failure} share for any model on any task, while smaller models keep this category below 17\% because their attempts often fail before reaching the verification stage. Together with F1, this pattern indicates that once models can reliably produce parsable Verus programs, the remaining challenge is not only surface syntax but also the proof reasoning needed to satisfy the verifier.

\subsection{Case Study}

\begin{figure}[t]
\centering
\begin{tcolorbox}[colback=white, colframe=black, boxrule=0.6pt, left=0pt, right=0pt, top=0pt, bottom=0pt, arc=0pt, outer arc=0pt]
\begin{lstlisting}[style=verusblock, firstnumber=1]
pub fn search_range(nums: Vec<i32>, target: i32) -> (result: Vec<i32>)
    requires
        0 <= nums.len() <= 100_000,
        forall |i: int| 0 <= i < nums.len()
            ==> -1_000_000_000 <= #[trigger] nums[i] <= 1_000_000_000,
        forall |i: int, j: int|
            0 <= i <= j < nums.len() ==> nums[i] <= nums[j],
        -1_000_000_000 <= target <= 1_000_000_000,
    ensures
        result.len() == 2,
        result[0] == -1i32 || result[0] >= 0,
        result[1] == -1i32 || result[1] >= 0,
        (forall |i: int| 0 <= i < nums.len() ==> nums[i] != target)
            ==> (result[0] == -1i32 && result[1] == -1i32),
        (result[0] == -1i32)
            ==> (forall |i: int| 0 <= i < nums.len() ==> nums[i] != target),
        result[0] >= 0 ==> (
            0 <= result[0] < nums.len() as i32
            && nums[result[0] as int] == target
            && (result[0] == 0i32
                || nums[result[0] as int - 1] < target)
        ),
        result[1] >= 0 ==> (
            0 <= result[1] < nums.len() as i32
            && nums[result[1] as int] == target
            && (result[1] == nums.len() as i32 - 1
                || nums[result[1] as int + 1] > target)
        ),
        (result[0] == -1i32) == (result[1] == -1i32),
        result[0] >= 0 ==> result[0] <= result[1],
\end{lstlisting}
\end{tcolorbox}
\caption{Ground-truth specification of \CodeIn{lc34}.}
\label{lst:lc34_gt}
\end{figure}

We illustrate the error patterns from Appendix~\ref{app:error_analysis} with \CodeIn{lc34} (Find First and Last Position of Element in Sorted Array) under the \CodeIn{end2end} setting. The problem asks for the first and last positions of the target value (\CodeIn{target}) in a non-decreasing array \CodeIn{nums}, returns \CodeIn{[-1,-1]} when \CodeIn{target} is absent, and requires an $O(\log n)$ algorithm. The ground-truth specification in Figure~\ref{lst:lc34_gt} constrains \CodeIn{nums} to be sorted with values in $[-10^9, 10^9]$, requires the output to be a length-two vector whose entries are either $-1$ or valid indices, and specifies that these entries identify the exact left and right boundaries of \CodeIn{target}. We compare the outputs of the two strongest models on this problem, Claude Opus 4.7 and GPT-5.5, to illustrate how different failure modes arise in the end-to-end setting.

\noindent\emph{\textbf{Claude Opus 4.7: verification succeeds against an incomplete specification.}}
Claude Opus 4.7 produces a correct two-pass binary search and proves it against the trivial postcondition \CodeIn{result.len() == 2} (Figure~\ref{lst:lc34_opus}). The program verifies in Verus and passes every functional test case, so verification and execution alone would not expose the specification error. However, both checks in the SpecGen metric reject this specification. The generated precondition omits the sortedness and value-range constraints required by the ground truth, so it fails the precondition equivalence check in Appendix~\ref{app:specgen_metric}. The generated postcondition does not relate \CodeIn{result} to \CodeIn{nums} or \CodeIn{target}, so the postcondition testing also fails. We therefore classify this attempt as \emph{Both Pre/Post Mismatch}. 

\begin{figure}[t]
\centering
\begin{tcolorbox}[colback=white, colframe=black, boxrule=0.6pt, left=0pt, right=0pt, top=0pt, bottom=0pt, arc=0pt, outer arc=0pt]
\begin{lstlisting}[style=verusblock, firstnumber=1]
pub fn search_range(nums: Vec<i32>, target: i32) -> (result: Vec<i32>)
    ensures
        result.len() == 2,   // ground-truth requires/ensures are dropped
{
    let n = nums.len();
    if n == 0 { return vec![-1, -1]; }
    let (mut lo, mut hi) = (0usize, n);
    while lo < hi
        invariant lo <= hi, hi <= n, n == nums.len(),
        decreases hi - lo,
    { /* lower-bound binary search */ }
    let first = lo;
    if first >= n || nums[first] != target { return vec![-1, -1]; }
    /* symmetric upper-bound pass; return [first, last] */
}
\end{lstlisting}
\end{tcolorbox}
\caption{Claude Opus 4.7 output for \CodeIn{lc34} under \CodeIn{end2end} (excerpt). Verus verifies the program and all positive tests pass, but the specification is reduced to \CodeIn{result.len() == 2}. The precondition equivalence check and the postcondition testing both reject the generated specification.}
\label{lst:lc34_opus}
\end{figure}

\noindent\emph{\textbf{GPT-5.5: stronger specification, but verification fails.}}
GPT-5.5 exhibits a different failure mode. It generates a non-trivial specification that includes the sortedness precondition and in-bounds, target-matching postconditions, factors the algorithm into \CodeIn{lower\_bound} and \CodeIn{upper\_bound} helpers, and adds detailed assertions for the loop invariants (Figure~\ref{lst:lc34_gpt55}). Despite the stronger specification, Verus rejects the program: the assertion \CodeIn{nums@[k] < target} on the lower-bound side and the symmetric assertion \CodeIn{nums@[k] <= target} on the upper-bound side both fail because the proof does not connect the sortedness property to the loop invariant strongly enough to discharge the inductive step. This case shows that once a model emits a parsable Verus program with a meaningful specification, the remaining challenge is proof construction rather than surface syntax, and even GPT-5.5 does not reliably discharge the obligations needed to verify an $O(\log n)$ binary search. 

\begin{figure}[t]
\centering
\begin{tcolorbox}[colback=white, colframe=black, boxrule=0.6pt, left=0pt, right=0pt, top=0pt, bottom=0pt, arc=0pt, outer arc=0pt]
\begin{lstlisting}[style=verusblock, firstnumber=1]
fn lower_bound(nums: &Vec<i32>, target: i32) -> (idx: usize)
    requires sorted(nums@), nums.len() <= 100000,
    ensures
        idx <= nums.len(),
        forall|k: int| 0 <= k < idx as int ==> nums@[k] < target,
        forall|k: int| idx as int <= k < nums@.len() ==> nums@[k] >= target,
{
    let (mut lo, mut hi) = (0usize, nums.len());
    while lo < hi
        invariant
            sorted(nums@), lo <= hi, hi <= nums.len(),
            forall|k: int| 0 <= k < lo as int ==> nums@[k] < target,
            forall|k: int| hi as int <= k < nums@.len()
                ==> nums@[k] >= target,
        decreases hi - lo,
    {
        let mid = lo + (hi - lo) / 2;
        if nums[mid] < target {
            assert(nums@[mid as int] < target);
            assert forall|k: int|
                0 <= k < mid as int + 1 ==> nums@[k] < target by {
                assert(nums@[k] < target); // <-- Verus: assertion failed
                assert(k <= mid as int);   // <-- Verus: assertion failed
            }
            lo = mid + 1;
        } else { hi = mid; }
    }
    lo
}
\end{lstlisting}
\end{tcolorbox}
\caption{GPT-5.5 output for \CodeIn{lc34} under \CodeIn{end2end} (excerpt). The generated specification is substantially stronger than the Claude Opus 4.7 specification, but Verus rejects two assertions inside the lower-bound loop because the proof does not connect the sortedness property to the loop invariant strongly enough to discharge the inductive step.}
\label{lst:lc34_gpt55}
\end{figure}

\subsection{Statistical Uncertainty and Paired Comparisons}
\label{app:cr_uncertainty}

We estimate uncertainty over benchmark problems using 20,000 bootstrap replicates of the 1,006 binary pass/fail outcomes. For a model with $s$ successes, we sample the replicate success count from $\mathrm{Binomial}(1006,s/1006)$ with random seed zero. Table~\ref{tab:cr_ci} reports the 2.5th and 97.5th percentiles as 95\% confidence intervals for all six tasks. These intervals capture variation from resampling problems; they do not include variation from repeated model calls.

\input{tables/bootstrap_ci_all_tasks}

For GPT-5.4 mini on \CodeIn{nl2code} and DeepSeek V4 Flash on \CodeIn{nl\_spec2code}, we compute the marginal intervals from the 844 and 649 successes reported in Table~\ref{tab:main_results}. The paired comparisons below use per-problem outcomes.

\textbf{Paired comparisons.} Since all models are evaluated on the same problems, we use two-sided exact McNemar tests~\cite{mcnemar1947note}. Table~\ref{tab:cr_paired_tests} reports selected comparisons. GPT-5.5 outperforms Claude Opus 4.7 on SpecGen ($p<10^{-50}$) and Claude Sonnet 4.6 on End2End ($p=6.7\times10^{-5}$). For ProofGen, we do not detect a difference between GPT-5.5 and Gemini 3.1 Pro ($p=0.45$), or between Gemini 3.1 Pro and Claude Sonnet 4.6 ($p=0.66$); these results do not imply that the models are equivalent. The reported $p$-values are unadjusted for multiple comparisons.

\input{tables/paired_mcnemar}

The multi-sample study in Figure~\ref{fig:proofgen_passk_repairk} measures the benefit of additional generations. On its 200-problem subset, ProofGen pass@1/pass@10 increase from 10.0\% to 21.5\% for Claude Sonnet 4.6, from 6.5\% to 21.5\% for Gemini 3.1 Pro, and from 2.5\% to 6.5\% for DeepSeek V4 Pro.

\input{sections/vericontest_lite}

\section{Limitations \& Discussion}
\label{app:limitations}

\textbf{Limitations.} First, \benchmark focuses on Rust programs with Verus. This choice targets a mainstream programming language and a modern verification framework, but it does not directly measure model performance in other verification systems such as Dafny~\cite{leino2010dafny} or Lean~\cite{moura2021lean}.

Second, \benchmark is limited to competitive-programming problems from LeetCode and Codeforces. These problems provide rich algorithmic reasoning and proof-heavy examples, but they do not cover all software domains, such as systems code, distributed protocols, or large repository-level verification. 

Finally, our construction pipeline is constrained by the current expressiveness of Verus and by the cost of manual review. Problems that require unsupported Rust features or proofs beyond the current capabilities of both agents and human experts are excluded. Future work can expand \benchmark to more languages, domains, and harder verification settings as verification tools and model capabilities improve.

\textbf{Specification validity.} Although the specifications in \benchmark undergo automated checks and manual review, we cannot guarantee that they are entirely free of errors or incompleteness. Despite our validation efforts, subtle specification issues may therefore remain in the benchmark.

\textbf{Data contamination.} Since LeetCode and Codeforces problems are public, some problem descriptions or solutions may appear in model pretraining corpora. This risk is most relevant to \CodeIn{nl2code}, where a model may benefit from exposure to the original description or a conventional solution. However, \benchmark also evaluates settings that require benchmark-specific Verus specifications and proofs. The strongest model drops from 97.51\% on \CodeIn{nl2code} to 77.44\% on \CodeIn{spec2code}, 57.16\% on SpecGen, 13.82\% on ProofGen, and 6.76\% on End2End. Thus, solution recall alone does not explain the main bottlenecks exposed by \benchmark.

We further measure verbatim overlap against the Rust subset of The Stack~\cite{kocetkov2022stack}, which contains 2,664,230 public source-code files. We lowercase and tokenize each artifact, treating identifiers, integer literals, and punctuation marks as tokens and discarding whitespace. For $n=10$, coverage is the fraction of token 10-grams that occur anywhere in the reference corpus. Table~\ref{tab:cr_contamination} reports overlap for specifications, executable code, and proofs. Specifications and proofs have lower mean coverage than executable code, and none of the 1,006 specifications or proofs exceeds 50\% coverage. Only eleven executable solutions (1.1\%) exceed 80\% coverage, and these are all short, canonical solutions for which common algorithmic patterns can also produce high overlap.

\input{tables/contamination_overlap}

\section{Prompts for Evaluation}
\label{app:evaluation_prompts}

Figures~\ref{fig:prompt_nl2code}--\ref{fig:prompt_proof_repair} show the prompt templates used for the six evaluation subtasks and proof repair. Each prompt follows the same structure: an instruction header, a one-shot example drawn from \CodeIn{task\_id\_809.rs} of Verus-Bench~\cite{yang2025autoverus}, task-specific input fields populated at evaluation time, an output field that fixes the response format, and a list of hard requirements. The example task is not part of \benchmark, so it does not leak benchmark content. Since the full worked examples are long, we replace their details with a short placeholder in the rendered templates while preserving the complete task instructions, input fields, output fields, and hard requirements.

\begin{figure}[H]
    \centering
    \begin{mdframed}[roundcorner=10pt]
    \input{prompts/nl2code.tex}
    \end{mdframed}
    \caption{Prompt for the \CodeIn{nl2code} setting.}
    \label{fig:prompt_nl2code}
\end{figure}

\begin{figure}[H]
    \centering
    \begin{mdframed}[roundcorner=10pt]
    \input{prompts/nl2spec.tex}
    \end{mdframed}
    \caption{Prompt for the \CodeIn{nl2spec} setting.}
    \label{fig:prompt_nl2spec}
\end{figure}

\begin{figure}[H]
    \centering
    \begin{mdframed}[roundcorner=10pt]
    \input{prompts/spec2code.tex}
    \end{mdframed}
    \caption{Prompt for the \CodeIn{spec2code} setting.}
    \label{fig:prompt_spec2code}
\end{figure}

\begin{figure}[H]
    \centering
    \begin{mdframed}[roundcorner=10pt]
    \input{prompts/nl_spec2code.tex}
    \end{mdframed}
    \caption{Prompt for the \CodeIn{nl\_spec2code} setting.}
    \label{fig:prompt_nl2spec2code}
\end{figure}

\begin{figure}[H]
    \centering
    \begin{mdframed}[roundcorner=10pt]
    \input{prompts/spec_code2proof.tex}
    \end{mdframed}
    \caption{Prompt for the \CodeIn{spec\_code2proof} setting.}
    \label{fig:prompt_spec_code2proof}
\end{figure}

\begin{figure}[H]
    \centering
    \begin{mdframed}[roundcorner=10pt]
    \input{prompts/end2end.tex}
    \end{mdframed}
    \caption{Prompt for the \CodeIn{end2end} setting.}
    \label{fig:prompt_end2end}
\end{figure}

\begin{figure}[H]
    \centering
    \begin{mdframed}[roundcorner=10pt]
    \input{prompts/proof_repair.tex}
    \end{mdframed}
    \caption{Prompt for proof repair.}
    \label{fig:prompt_proof_repair}
\end{figure}

\clearpage
\endgroup

%% file: tables/construction_attrition.tex
\begin{table}[H]
\centering
\captionsetup{skip=6pt}
\caption{Construction attrition by source. Screening exclusion rates use the screened count as denominator; verification failure rates use the attempted count.}
\label{tab:cr_attrition}
\small
\scalebox{0.88}{
\begin{tabular}{@{}lrr@{}}
\toprule
\textbf{Stage} & \textbf{LeetCode} & \textbf{Codeforces} \\
\midrule
Eligible candidate pool & 3,106 & 6,062 \\
Screened & 3,106 & 548 \\
Excluded during screening & 2,153 (69.3\%) & 141 (25.7\%) \\
Attempted & 953 & 407 \\
Failed verification within construction budget & 231 (24.2\%) & 123 (30.2\%) \\
Accepted & 722 & 284 \\
\bottomrule
\end{tabular}
}
\end{table}

%% file: tables/construction_exclusions.tex
\begin{table}[H]
\centering
\captionsetup{skip=6pt}
\caption{Screening exclusions and verification failures. Percentages are calculated within each group.}
\label{tab:cr_exclusions}
\small
\begin{tabular}{@{}llrr@{}}
\toprule
\textbf{Stage} & \textbf{Category} & \textbf{Count} & \textbf{Share (\%)} \\
\midrule
Screening & Unsupported features & 1,019 & 44.4 \\
& Complex data structures & 897 & 39.1 \\
& LeetCode hidden structures & 178 & 7.8 \\
& SQL, shell, or concurrency tasks & 200 & 8.7 \\
\midrule
Verification & Greedy/DP optimality proofs & 159 & 44.9 \\
& Complex data structures & 95 & 26.8 \\
& Mathematics and number theory & 82 & 23.2 \\
& Matrix reasoning & 18 & 5.1 \\
\bottomrule
\end{tabular}
\end{table}

%% file: tables/filter_distribution.tex
\begin{table}[H]
\centering
\captionsetup{skip=6pt}
\caption{Recorded difficulty and tag distributions before and after filtering (\%). Only top five tags are shown here.}
\label{tab:cr_filter_distribution}
\small
\begin{tabular}{@{}llrr@{}}
\toprule
\textbf{Source} & \textbf{Difficulty or tag} & \textbf{Before} & \textbf{After} \\
\midrule
LeetCode & Easy & 25.8 & 40.2 \\
& Medium & 50.5 & 43.5 \\
& Hard & 23.7 & 16.3 \\
\cmidrule(l){2-4}
& Array & 58.5 & 72.6 \\
& Math & 18.6 & 28.3 \\
& Dynamic programming & 18.3 & 25.8 \\
& Sorting & 14.2 & 17.3 \\
& Greedy & 13.2 & 17.0 \\
\midrule
Codeforces & 800--1000 & 29.3 & 42.5 \\
& 1100--1300 & 22.1 & 28.3 \\
& 1400--1900 & 48.6 & 29.2 \\
\cmidrule(l){2-4}
& Greedy & 38.2 & 39.1 \\
& Implementation & 37.7 & 39.2 \\
& Math & 33.9 & 39.5 \\
& Brute force & 19.7 & 19.5 \\
& Constructive algorithms & 19.6 & 20.2 \\
\bottomrule
\end{tabular}
\end{table}

%% file: tables/llms.tex
\begin{table*}[t]
\centering
\caption{Detailed Configurations and Costs for the Evaluated LLMs}
\label{tab:llms}
\Large
\scalebox{0.56}{
\begin{tabular}{llllcc}
\toprule
\textbf{Vendor} & \textbf{Model Name} & \textbf{Checkpoint} & \textbf{Type} & \begin{tabular}[c]{@{}l@{}}\textbf{Price (\$/1M tokens)}\\\textbf{(Input/Output)}\end{tabular} & \textbf{Cost} \\
\midrule
\multirow{2}{*}{OpenAI} & GPT-5.5 \cite{gpt_5_5} & gpt-5.5-2026-04-23 & API & \$5.00 / \$30.00 & \$987.31 \\
                        & GPT-5.4 mini \cite{gpt_5_4_mini} & gpt-5.4-mini-2026-03-17 & API & \$0.75 / \$4.50 & \$94.55 \\
\midrule
\multirow{2}{*}{Anthropic} & Claude Opus 4.7 \cite{claude_opus_4_7} & claude-opus-4-7 & API & \$5.00 / \$25.00 & \$542.63 \\
                           & Claude Sonnet 4.6 \cite{claude_sonnet_4_6} & claude-sonnet-4-6 & API & \$3.00 / \$15.00 & \$393.11 \\
\midrule            
\multirow{2}{*}{Google} & Gemini 3.1 Pro \cite{gemini_3_1_pro} & gemini-3.1-pro-preview & API & \$2.00 / \$12.00 & \$450.21 \\
                        & Gemini 3 Flash \cite{gemini_3_flash} & gemini-3-flash-preview & API & \$0.25 / \$1.50 & \$31.88 \\
\midrule  
\multirow{2}{*}{DeepSeek} & DeepSeek V4 Pro \cite{deepseekai2026deepseekv4} & deepseek-v4-pro & API & \$1.74 / \$3.48 & \$124.34 \\
                            & DeepSeek V4 Flash \cite{deepseekai2026deepseekv4} & deepseek-v4-flash & API & \$0.14 / \$0.28 & \$7.81 \\
\midrule  
Alibaba & Qwen 3.6 \cite{qwen3.6-27b} & Qwen3.6-27B & GPU & -- & -- \\
\midrule
Zhipu AI & GLM-4.7-Flash \cite{5team2025glm45agenticreasoningcoding} & GLM-4.7-Flash & GPU & -- & -- \\
\bottomrule
\end{tabular}
}
\end{table*}

%% file: tables/difficulty_stratified.tex
\begin{table}[H]
\centering
\captionsetup{skip=6pt}
\caption{Pass@1 by source and difficulty.}
\label{tab:cr_difficulty}
\small
\setlength{\tabcolsep}{4pt}
\begin{tabular}{@{}llrrrrrr@{}}
\toprule
& & \multicolumn{3}{c}{\textbf{LeetCode}} & \multicolumn{3}{c}{\textbf{Codeforces}} \\
\cmidrule(lr){3-5}\cmidrule(l){6-8}
\textbf{Model} & \textbf{Task} & \textbf{Easy} & \textbf{Medium} & \textbf{Hard} & \textbf{800--1000} & \textbf{1100--1300} & \textbf{1400--1900} \\
\midrule
GPT-5.5 & CodeGen & 95.8 & 98.3 & 98.4 & 98.1 & 98.6 & 100.0 \\
 & SpecGen & 66.6 & 54.6 & 57.4 & 51.3 & 51.4 & 33.3 \\
 & ProofGen & 16.7 & 6.6 & 6.6 & 30.4 & 6.9 & 3.7 \\
 & End2End & 10.0 & 4.6 & 1.6 & 8.9 & 2.8 & 1.9 \\
\midrule
Claude Sonnet 4.6 & CodeGen & 98.9 & 95.0 & 91.8 & 91.8 & 77.8 & 83.3 \\
 & SpecGen & 25.9 & 16.2 & 8.2 & 14.6 & 2.8 & 3.7 \\
 & ProofGen & 15.0 & 8.3 & 4.9 & 25.9 & 16.7 & 1.9 \\
 & End2End & 4.5 & 2.3 & 1.6 & 6.3 & 0.0 & 1.9 \\
\bottomrule
\end{tabular}
\end{table}

%% file: tables/bootstrap_ci_all_tasks.tex
\begin{table*}[t]
\centering
\caption{Pass@1 with 95\% problem-bootstrap confidence intervals across all six evaluation tasks. Each cell reports the point estimate followed by the confidence interval in brackets.}
\label{tab:cr_ci}
\small
\scalebox{0.73}{
\begin{tabular}{l|c|ccc|c|c}
\toprule
\textbf{Model} &
\multicolumn{1}{c}{\textbf{SpecGen}} &
\multicolumn{3}{c}{\textbf{CodeGen}} &
\multicolumn{1}{c}{\textbf{ProofGen}} &
\multicolumn{1}{c}{\textbf{End-to-End}} \\
\cmidrule(lr){2-2}
\cmidrule(lr){3-5}
\cmidrule(lr){6-6}
\cmidrule(lr){7-7}
& nl2spec
& nl2code
& spec2code
& nl\_spec2code
& spec\_code2proof
& end2end \\
\midrule
GPT-5.5 & 57.16 [54.2, 60.2] & 97.51 [96.5, 98.4] & 77.44 [74.9, 80.0] & 84.49 [82.3, 86.7] & 13.82 [11.7, 16.0] & 6.76 [5.3, 8.3] \\
GPT-5.4 mini & 7.85 [6.3, 9.5] & 83.90 [81.6, 86.2] & 54.57 [51.5, 57.7] & 63.02 [60.0, 66.0] & 6.36 [4.9, 8.0] & 1.49 [0.8, 2.3] \\
\midrule
Claude Opus 4.7 & 27.24 [24.6, 30.0] & 95.43 [94.1, 96.7] & 77.24 [74.7, 79.8] & 80.62 [78.1, 83.0] & 12.52 [10.5, 14.6] & 3.58 [2.5, 4.8] \\
Claude Sonnet 4.6 & 17.30 [15.0, 19.7] & 93.84 [92.2, 95.3] & 82.01 [79.6, 84.4] & 83.80 [81.4, 86.0] & 13.52 [11.4, 15.7] & 3.48 [2.4, 4.7] \\
\midrule
Gemini 3.1 Pro & 22.56 [20.1, 25.1] & 94.14 [92.6, 95.5] & 80.62 [78.2, 83.0] & 89.86 [88.0, 91.7] & 12.92 [10.8, 15.0] & 3.68 [2.6, 4.9] \\
Gemini 3 Flash & 22.66 [20.1, 25.3] & 53.08 [50.0, 56.2] & 32.50 [29.6, 35.5] & 36.58 [33.7, 39.6] & 6.16 [4.7, 7.7] & 1.29 [0.6, 2.0] \\
\midrule
DeepSeek V4 Pro & 12.13 [10.1, 14.2] & 92.05 [90.4, 93.6] & 66.80 [63.8, 69.7] & 73.06 [70.4, 75.7] & 5.47 [4.1, 7.0] & 1.29 [0.7, 2.1] \\
DeepSeek V4 Flash & 9.74 [8.0, 11.6] & 82.21 [79.8, 84.5] & 64.31 [61.3, 67.3] & 64.51 [61.6, 67.5] & 6.36 [4.9, 8.0] & 1.19 [0.6, 1.9] \\
\midrule
Qwen 3.6 & 0.40 [0.1, 0.8] & 82.50 [80.1, 84.8] & 43.94 [40.9, 47.0] & 51.59 [48.6, 54.7] & 4.57 [3.3, 5.9] & 0.20 [0.0, 0.5] \\
\midrule
GLM-4.7-Flash & 3.68 [2.6, 4.9] & 43.34 [40.3, 46.4] & 17.10 [14.8, 19.5] & 22.17 [19.6, 24.8] & 5.27 [4.0, 6.7] & 0.70 [0.2, 1.3] \\
\bottomrule
\end{tabular}
}
\end{table*}

%% file: tables/paired_mcnemar.tex
\begin{table}[H]
\centering
\captionsetup{skip=6pt}
\caption{Selected two-sided exact paired McNemar tests.}
\label{tab:cr_paired_tests}
\small
\begin{tabular}{@{}lll@{}}
\toprule
\textbf{Task} & \textbf{Model pair} & \textbf{$p$-value} \\
\midrule
SpecGen & GPT-5.5 vs. Claude Opus 4.7 & $<10^{-50}$ \\
End2End & GPT-5.5 vs. Claude Sonnet 4.6 & $6.7\times10^{-5}$ \\
ProofGen & GPT-5.5 vs. Gemini 3.1 Pro & $0.45$ \\
ProofGen & Gemini 3.1 Pro vs. Claude Sonnet 4.6 & $0.66$ \\
\bottomrule
\end{tabular}
\end{table}

%% file: sections/vericontest_lite.tex
\section{\benchmark Lite}
\label{app:vericontest_lite}

\subsection{Motivation and Scope}

Evaluating all \benchsize problems becomes costly when comparing multiple models, testing prompting strategies, or running agents that make repeated generation and proof-repair calls. We therefore introduce \emph{\benchmark Lite}, a subset of 300 problems for lightweight evaluation. Lite contains 215 LeetCode and 85 Codeforces problems and supports the same six task settings as the full benchmark. It retains the selected problems' descriptions, specifications, code, proofs, and test suites, together with the evaluation criteria in Section~\ref{sec:evaluation_metrics}. 

\subsection{Subset Construction}

\textbf{Eligibility filters.} Starting from the \benchsize problems, we exclude problems whose descriptions contain images or hyperlinks, refer to another problem, require a problem-specific output checker, or whose reference solution exceeds 0.5\,s on any test case under the 1\,s per-case limit. The first two filters favor self-contained text inputs. The third restricts Lite to outputs evaluated by exact comparison, simplifying reproduction across evaluation harnesses. The final filter provides a runtime margin to reduce sensitivity to hardware and system load. These filters leave 914 eligible problems.

\textbf{Stratified sampling.} We allocate 300 places in proportion to the \emph{full benchmark's} source and difficulty counts, using largest-remainder rounding first across sources and then across difficulty groups within each source. Table~\ref{tab:lite_composition} lists the resulting allocation. Within each group, we sort all full-benchmark problems by reference proof size and partition the sorted list into one interval per allocated place. We draw one target problem uniformly from each interval using the fixed seed. If a target is ineligible or already selected, we choose an unselected eligible problem in the same group with the closest proof size. This procedure preserves the full benchmark's allocation even when filtering changes the eligible pool.

\input{tables/lite_composition}

\textbf{Proof-size coverage.} The resulting subset closely follows the reference proof-size distribution. For LeetCode, the median is 76 lines in Lite versus 77 in the full benchmark, and the 90th percentile is 312 versus 326 lines. For Codeforces, the corresponding medians are 108 and 110 lines, with a 90th percentile of 351 lines in both splits. 

\subsection{Model Performance}

Table~\ref{tab:lite_results} reports pass@1 for the ten evaluated models across all six settings. The model configurations and prompting remain those described in Appendix~\ref{app:evaluation_details}.

\input{tables/lite_results}

\textbf{Lite retains the main capability gaps.} GPT-5.5 achieves 97.00\% on \CodeIn{nl2code}, but only 57.00\% on SpecGen, 14.67\% on ProofGen, and 6.33\% on End2End. GPT-5.5 leads SpecGen, natural-language CodeGen, and End2End, and ties Claude Opus 4.7 on ProofGen. Gemini 3.1 Pro leads both specification-based CodeGen settings. No model exceeds 14.67\% on ProofGen or 6.33\% on End2End, preserving the full benchmark's finding that specification and proof generation remain major obstacles to verifiable code generation.

\textbf{Agreement with the full benchmark.} Across all 60 model--task combinations, the largest absolute difference between Lite and full-benchmark pass@1 is 2.92 percentage points. Rankings are therefore similar but not identical. For example, Claude Opus 4.7 and GPT-5.5 reverse order on \CodeIn{nl\_spec2code}: Lite scores are 83.33\% and 82.67\%, whereas full-benchmark scores are 80.62\% and 84.49\%. Each Lite problem changes pass@1 by 0.33 percentage points, so small score differences warrant caution. These results support Lite for frequent evaluation and iteration, with the full benchmark providing a broader assessment for close comparisons.

%% file: tables/lite_composition.tex
\begin{table}[t]
\centering
\caption{Source and difficulty composition of \benchmark Lite.}
\label{tab:lite_composition}
\small
\begin{tabular}{llrr}
\toprule
\textbf{Source} & \textbf{Difficulty} & \textbf{Full} & \textbf{Lite} \\
\midrule
\multirow{3}{*}{LeetCode} & Easy & 359 & 107 \\
& Medium & 302 & 90 \\
& Hard & 61 & 18 \\
\midrule
\multirow{5}{*}{Codeforces} & Rating 800 & 118 & 35 \\
& Rating 900--1000 & 40 & 12 \\
& Rating 1100--1200 & 49 & 15 \\
& Rating 1300--1400 & 40 & 12 \\
& Rating 1500+ & 37 & 11 \\
\midrule
\multicolumn{2}{l}{Total} & 1,006 & 300 \\
\bottomrule
\end{tabular}
\end{table}

%% file: tables/lite_results.tex
\begin{table*}[t]
\centering
\caption{Main results of pass@1 accuracy across six evaluation tasks on \benchmark Lite. The \textbf{\highlight{Green!15}{highest}} and \highlight{Cyan!15}{second-highest} accuracies are highlighted.}
\label{tab:lite_results}
\small
\scalebox{0.88}{
\begin{tabular}{l|c|ccc|c|c}
\toprule
\textbf{Model} &
\multicolumn{1}{c}{\textbf{SpecGen}} &
\multicolumn{3}{c}{\textbf{CodeGen}} &
\multicolumn{1}{c}{\textbf{ProofGen}} &
\multicolumn{1}{c}{\textbf{End-to-End}} \\
\cmidrule(lr){2-2}
\cmidrule(lr){3-5}
\cmidrule(lr){6-6}
\cmidrule(lr){7-7}
& nl2spec
& nl2code
& spec2code
& nl\_spec2code
& spec\_code2proof
& end2end \\
\midrule
GPT-5.5 & \textbf{\highlight{Green!15}{57.00}} & \textbf{\highlight{Green!15}{97.00}} & 75.00 & 82.67 & \textbf{\highlight{Green!15}{14.67}} & \textbf{\highlight{Green!15}{6.33}} \\
GPT-5.4 mini & 7.00 & 84.67 & 53.00 & 61.00 & 6.33 & 0.67 \\
\midrule
Claude Opus 4.7 & \highlight{Cyan!15}{29.33} & \highlight{Cyan!15}{96.33} & 78.33 & 83.33 & \textbf{\highlight{Green!15}{14.67}} & 2.00 \\
Claude Sonnet 4.6 & 16.33 & 93.67 & \highlight{Cyan!15}{81.00} & \highlight{Cyan!15}{86.33} & 12.00 & \highlight{Cyan!15}{2.67} \\
\midrule
Gemini 3.1 Pro & 22.33 & 92.33 & \textbf{\highlight{Green!15}{82.67}} & \textbf{\highlight{Green!15}{87.67}} & \highlight{Cyan!15}{13.67} & \highlight{Cyan!15}{2.67} \\
Gemini 3 Flash & 21.33 & 56.00 & 30.67 & 37.67 & 6.33 & 1.00 \\
\midrule
DeepSeek V4 Pro & 12.33 & 94.33 & 65.00 & 73.67 & 5.33 & 1.00 \\
DeepSeek V4 Flash & 9.33 & 85.00 & 62.00 & 62.00 & 5.33 & 0.33 \\
\midrule
Qwen 3.6 & 0.00 & 83.67 & 45.33 & 51.33 & 4.67 & 0.00 \\
\midrule
GLM-4.7-Flash & 3.67 & 42.67 & 15.00 & 19.67 & 5.33 & 0.33 \\
\bottomrule
\end{tabular}
}
\end{table*}

%% file: tables/contamination_overlap.tex
\begin{table}[!htbp]
\centering
\captionsetup{skip=6pt}
\caption{Reported 10-gram overlap with The Stack's Rust subset. The max verbatim span is the largest contiguous chain of corpus-matched token $n$-grams. The proof span excludes a 141-token comment divider (a line of dashes) in \CodeIn{cf1506D}.}
\label{tab:cr_contamination}
\scalebox{0.90}{
\begin{tabular}{@{}lrrrr@{}}
\toprule
\textbf{Artifact} & \textbf{Mean coverage} & \textbf{Max verbatim span} & \textbf{Coverage $>50\%$} & \textbf{Coverage $>80\%$}\\
\midrule
Specification & 6.1\% & 44 tokens & 0 / 1,006 & 0 / 1,006 \\
Executable code & 28.1\% & 96 tokens & 83 / 1,006 & 11 / 1,006 \\
Proof & 0.7\% & 37 tokens & 0 / 1,006 & 0 / 1,006 \\
\bottomrule
\end{tabular}
}
\end{table}

%% file: prompts/nl2code.tex
\begin{lstlisting}[style=promptstyle]
# Instruction

You are an expert in Rust. You will be provided with a natural language problem description. Your task is to generate the Rust code from the problem description.

# Example

[A one-shot Verus example here. We omit the example details in the paper for space.]

# Input Field

```
{{problem_description}}
```

# Output Field

Output a Rust program. Your answer should only contain a Rust program (you must quote the program in **```rust** for further automation):

```rust
{{Rust program}}
```

**CRITICAL REQUIREMENTS:**

- **The code should be written in a Verus-verifiable format.**

- **Ensure the code is efficient and avoids timeouts.**

- **Do not include explanations, analysis, markdown outside the Rust code fence, or multiple alternative programs.**
\end{lstlisting}

%% file: prompts/nl2spec.tex
\begin{lstlisting}[style=promptstyle]
# Instruction

You are an expert in Verus. You will be provided with a natural language problem description. Your task is to generate the Verus specification from the problem description.

# Example

[A one-shot Verus example here. We omit the example details in the paper for space.]

# Input Field

```
{{problem_description}}
```

# Output Field

Output a Verus specification. Your answer should only contain a Verus program with only the specification as follows (you must quote the program in **```rust** for further automation):

```rust
{{Verus program containing only the specification}}
```

**CRITICAL REQUIREMENTS:**

- **Generate only Verus specification. Do not include any code or proofs.**

- **Do not include explanations, analysis, markdown outside the Rust code fence, or multiple alternative programs.**
\end{lstlisting}
    

%% file: prompts/spec2code.tex
\begin{lstlisting}[style=promptstyle]
# Instruction

You are an expert in Verus. You will be provided with a Verus specification. Your task is to generate the Rust code that aligns with the specification.

# Example

[A one-shot Verus example here. We omit the example details in the paper for space.]

# Input Field

```rust
{{verus_spec}}
```

# Output Field

Output a Verus program with the specification. Your answer should only contain a Verus program as follows (you must quote the program in **```rust** for further automation):

```rust
{{Verus program with the specification}}
```

**CRITICAL REQUIREMENTS:**

- **Generate only Rust code. Do not include any proofs.**

- **Ensure the code is efficient and avoids timeouts.**

- **The code should be written in a Verus-verifiable format.**

- **Do not include explanations, analysis, markdown outside the Rust code fence, or multiple alternative programs.**
\end{lstlisting}

%% file: prompts/nl_spec2code.tex
\begin{lstlisting}[style=promptstyle]
# Instruction

You are an expert in Verus. You will be provided with a natural language problem description and a Verus specification. Your task is to generate the Rust code from them.

# Example

[A one-shot Verus example here. We omit the example details in the paper for space.]

# Input Field

## Problem Description

```
{{problem_description}}
```

## Verus Specification

```rust
{{verus_spec}}
```

# Output Field

Output a Verus program. Your answer should only contain a Verus program along with the specification as follows (you must quote the program in **```rust** for further automation):

```rust
{{Verus program}}
```

**CRITICAL REQUIREMENTS:**

- **Generate only Rust code. Do not include any Verus proofs.**

- **The code should be written in a Verus-verifiable format.**

- **Do not include explanations, analysis, markdown outside the Rust code fence, or multiple alternative programs.**

- **Ensure the code is efficient and avoids timeouts.**
\end{lstlisting}
    

%% file: prompts/spec_code2proof.tex
\begin{lstlisting}[style=promptstyle]
# Instruction

You are an expert in Verus. You will be provided with a Verus specification along with the executable Rust code. Your task is to generate the Verus proofs to verify this program.

# Example

[A one-shot Verus example here. We omit the example details in the paper for space.]

# Input Field

```rust
{{verus_program}}
```

# Output Field

Output a verified Verus program with proofs. Your answer should only contain a verified Verus program as follows (you must quote the program in **```rust** for further automation):

```rust
{{Verus program}}
```

**CRITICAL REQUIREMENTS:**

- **Preserve the original executable Rust code and specification.**

- **Do not include explanations, analysis, markdown outside the Rust code fence, or multiple alternative programs.**
\end{lstlisting}
    

%% file: prompts/end2end.tex
\begin{lstlisting}[style=promptstyle]
# Instruction

You are an expert in Verus. You will be provided with a natural language problem description. Your task is to generate a verified Verus program from the problem description, including the Verus specification, executable Rust code, and the corresponding Verus proofs.

# Example

[A one-shot Verus example here. We omit the example details in the paper for space.]

# Input Field

```
{{problem_description}}
```

# Output Field

Output a verified Verus program. Your answer should only contain a Verus program (you must quote the program in **```rust** for further automation):

```rust
{{Verified Verus program}}
```

**CRITICAL REQUIREMENTS:**

- **Ensure the code is efficient and avoids timeouts.**

- **Generate the Verus specification, executable Rust code, and the corresponding Verus proofs together.**

- **Do not include explanations, analysis, markdown outside the Rust code fence, or multiple alternative programs.**
\end{lstlisting}
    

%% file: prompts/proof_repair.tex
\begin{lstlisting}[style=promptstyle]
# Instruction

You are an expert in Verus. You will be provided with an incorrect Verus program and the error messages emitted by Verus. Your task is to repair the proof so that the program verifies.

You should preserve the original executable Rust code and the specification. Only change the proofs of the program.

# Example

[A one-shot Verus example here. We omit the example details in the paper for space.]

# Input Field

## Incorrect Verus Program

```rust
{{verus_program}}
```

## Verus Error

```text
{{error_message}}
```

# Output Field

Output the repaired Verus program. Your answer should only contain the repaired Verus program as follows (you must quote the program in **```rust** for further automation):

```rust
{{Repaired Verus program}}
```

**CRITICAL REQUIREMENTS:**

- **Preserve the original executable Rust code and specification.**

- **Do not include explanations, analysis, markdown outside the Rust code fence, or multiple alternative programs.**
\end{lstlisting}
    